\documentclass[pre,twocolumn,10pt,aps,longbibliography,nofootinbib]{revtex4-1}

 \usepackage{bm}
 \usepackage{amsmath}
 \usepackage{amssymb}
 \usepackage{dsfont}
 \usepackage{latexsym}
 \usepackage{amsfonts}
 \usepackage{epsfig}
 \usepackage{epstopdf}
 \usepackage{color}
 \definecolor{darkblue}{rgb}{0,0,.5}
 \usepackage[linktocpage, colorlinks=true, linkcolor=darkblue, citecolor=darkblue]{hyperref}
 \usepackage[all]{hypcap}

\newcommand{\C}[1]{{\cal{#1}}}
\newcommand{\bb}[1]{\textbf{#1}}

\newcommand{\braketmat}[3]{\left<#1\left|#2\right|#3\right>}
\newcommand{\abs}[1]{{\left| #1 \right|}}

\newcommand{\rl}[0]{{\rangle\langle}}

\begin{document}

\title{Nonequilibrium thermodynamics in the strong coupling and non-Markovian regime based on a
reaction coordinate mapping}

\author{Philipp Strasberg${}^1$}
\author{Gernot Schaller${}^1$}
\author{Neill Lambert${}^2$}
\author{Tobias Brandes${}^1$}
\affiliation{$^1$ {Institut f\"ur Theoretische Physik, Technische Universit\"at Berlin, Hardenbergstr. 36, D-10623 Berlin, Germany} \\
$^2$ CEMS, RIKEN, Saitama, 351-0198, Japan}

\begin{abstract}
 We propose a method to study the thermodynamic behaviour of small systems beyond the weak coupling and Markovian
 approximation, which is different in spirit from conventional approaches. 
 The idea is to redefine the system and environment such that the effective, redefined system is again coupled 
 weakly to Markovian residual baths and thus, allows to derive a consistent thermodynamic framework for this new 
 system-environment partition. To achieve this goal we make use of the reaction coordinate mapping, which is a general 
 method
 in the sense that it can be applied to an arbitrary
 (quantum or classical and even time-dependent) system coupled linearly to an arbitrary number of harmonic oscillator
 reservoirs. The core of the method relies on an appropriate identification of a part of the environment (the reaction
 coordinate), which is subsequently included as a part of the system. 
 We demonstrate the power of this concept by showing that non-Markovian effects can significantly enhance the
 steady state efficiency of a three-level-maser heat engine, even in the regime of weak system-bath coupling. 
 Furthermore, we show for a single electron transistor coupled to vibrations that our method allows one to justify 
 master equations derived in a polaron transformed reference frame.
\end{abstract}


\maketitle

\section{Introduction}

Classical thermodynamics is a weak coupling theory in the sense that boundary or surface terms of the system
are negligible compared to its bulk or volume properties. This becomes particularly apparent in Maxwell's colloquial
description of the zeroth law of thermodynamics: ``All heat is of the same kind''\cite{MaxwellBook1871}. By this
statement he meant that the laws governing the transformation of heat are independent of how we put two different
systems into contact -- a conclusion which obviously holds only if the influence of this contact can be neglected.
Other implications of the weak coupling approximation are, for instance, the extensiveness of internal energy or entropy 
if we scale the volume of a system~\cite{LiebYngvasonPR1999}.

While the weak coupling approximation can be well justified for macroscopic systems due to simple geometric arguments
(the surface to volume ratio usually decreases with increasing volume), it is harder to justify in the opposite limit
when the volume of the system becomes very small. This, however, is the regime where quantum and stochastic effects
dominate. Then, in order to link microscopic theory with thermodynamics, one usually starts with a Hamiltonian of
the form
\begin{equation}\label{eq system bath Hamiltonian}
 H = H_S + H_E + H_I
\end{equation}
where $H_S$ ($H_E$) is the Hamiltonian of the system (environment) and $H_I$ describes their interaction (the
``contact''). Assuming that the coupling $H_I$ is small, one then performs a perturbative expansion up to second order
in $H_I$ which (under the additional assumption that the environment is memory-less) yields a closed and
Markovian evolution equation for the system density matrix $\rho_S$, known as a (quantum) master equation (ME). We note
that similar assumptions are needed to derive (classical) Fokker-Planck or Langevin equations. MEs derived this way can
then be shown to have a transparent thermodynamic interpretation~\cite{SpohnJMP1978, SpohnLebowitzAdvChemPhys1979,
AlickiJPA1979} and they provide the work horse for the field of quantum and stochastic thermodynamics, see
Refs.~\cite{EspositoHarbolaMukamelRMP2009, CampisiHaenggiTalknerRMP2011, KosloffEntropy2013, SchallerBook2014,
GelbwaserKlimovskyNiedenzuKurizkiAdv2015} for recent reviews.

However, many interesting physical effects cannot be captured with such a ME approach and thus, quantum and stochastic
thermodynamics is still restricted to a small regime of applicability. Consequently, many groups have started to
look at thermodynamics in the strong coupling and non-Markovian regime~\cite{SpeckSeifertJSM2007,
EspositoLindenbergVandenBroeckNJP2010, AnkerholdPekolaPRB2014, JarzynskiJSM2004, CampisiTalknerHaenggiPRL2009, 
GallegoRieraEisertNJP2014, 
SeifertPRL2016, KawamotoHatanoPRE2011, LeggioEtAlPRE2013, VinjanampathyModiArXiv2014, VinjanampathyModiPRA2015,
EspositoOchoaGalperinPRL2015, ToppBrandesSchallerEPL2015, EspositoOchoaGalperinPRB2015, BruchEtAlPRB2016,
KimMahlerPRE2010, AdamietzIngoldWeissEPJB2014, CarregaEtAlNJP2015, PhilbinAndersJPA2016, SegalPRB2006, 
NicolinSegalPRB2011, SaitoKatoPRL2013, SchallerEtAlNJP2013, KrauseEtAlJCP2015, WangRenCaoSciRep2015, 
GelbwaserKlimovskyAspuruGuzikJPCL2015, BylickaEtAlArXiv2015}. 
Though these works present important 
theoretical cornerstones, they are still far away from providing a satisfactory extension of thermodynamics beyond the 
weak coupling limit. In particular, 
if one wishes to address the performance of a steadily working heat engine, the 
general results derived in Refs.~\cite{SpeckSeifertJSM2007, EspositoLindenbergVandenBroeckNJP2010, AnkerholdPekolaPRB2014,
JarzynskiJSM2004, CampisiTalknerHaenggiPRL2009, GallegoRieraEisertNJP2014, SeifertPRL2016, KawamotoHatanoPRE2011,
LeggioEtAlPRE2013, VinjanampathyModiArXiv2014, VinjanampathyModiPRA2015} are not of great help because they either
focus on integrated changes of thermodynamic values (e.g., the total heat exchanged in a finite time instead of the
\emph{rate} of heat exchange) and additionally rely on an initially
decorrelated system-environment state~\cite{SpeckSeifertJSM2007, EspositoLindenbergVandenBroeckNJP2010,
AnkerholdPekolaPRB2014} and/or coupling only to a single thermal reservoir~\cite{SpeckSeifertJSM2007,
AnkerholdPekolaPRB2014, JarzynskiJSM2004, CampisiTalknerHaenggiPRL2009, GallegoRieraEisertNJP2014, SeifertPRL2016};
or they remain very formal~\cite{KawamotoHatanoPRE2011, LeggioEtAlPRE2013, VinjanampathyModiArXiv2014,
VinjanampathyModiPRA2015}. Furthermore, model-specific studies are either based on simple or exactly solvable models
from the field of quantum transport~\cite{EspositoOchoaGalperinPRL2015, ToppBrandesSchallerEPL2015,
EspositoOchoaGalperinPRB2015, BruchEtAlPRB2016} and  quantum Brownian motion~\cite{KimMahlerPRE2010, 
AdamietzIngoldWeissEPJB2014, CarregaEtAlNJP2015, PhilbinAndersJPA2016}, or spin-boson 
models ~\cite{SegalPRB2006, NicolinSegalPRB2011,
SaitoKatoPRL2013} often in combination with specific transformations applicable only to special Hamiltonians (polaron
transformations)~\cite{NicolinSegalPRB2011, SchallerEtAlNJP2013, KrauseEtAlJCP2015, WangRenCaoSciRep2015,
GelbwaserKlimovskyAspuruGuzikJPCL2015}; or the investigations are restricted to
numerical studies~\cite{BylickaEtAlArXiv2015}. 

The goal of this paper is to close the gap between the general results, which are often hard to apply in practice, and
studies restricted to overly-specific models. Here, we propose a framework which allows to carry over all
concepts known from the weak coupling regime of thermodynamics to the strong coupling and non-Markovian regime.\footnote{We wish 
to remark that, 
though often correlated, the concepts of strong coupling and non-Markovianity can be defined separately. Especially, 
a system can be strongly coupled to an environment but behave purely Markovian and, vice versa, it can be coupled very 
weakly but behave strongly non-Markovian. } Our framework is general in the sense that
it can be applied to an \emph{arbitrary} (quantum or classical\footnote{The notation we are using is adapted to the
quantum mechanical situation but all transformations carry over to the classical situation, too. }
and even driven) system coupled linearly to an arbitrary number of harmonic oscillator heat baths. Thus, apart from 
not being able to treat, e.g., fermionic reservoirs at the moment, we capture
many relevant situations encountered in the study of small-scale engines.

The general idea is to give up the system-environment partition as it is dictated by the microscopic 
Hamiltonian~(\ref{eq system bath Hamiltonian}). Instead, we define a new ``supersystem'' which includes this part of the 
environment which is responsible for strong coupling and non-Markovian effects. By construction the resulting supersystem 
would be 
coupled weakly to Markovian residual baths and can be treated within the standard framework of quantum or stochastic 
thermodynamics. More specifically,
to achieve this idea, we will identify a collective degree of freedom in the reservoir which is then 
incorporated into the description of the system. This collective degree of freedom is known as a \emph{reaction 
coordinate} (RC)~\cite{GargOnuchicAmbegaokarJCP1985}, which will capture non-Markovian and strong coupling effects. For 
this ``supersystem'' (original system plus RC) we will derive the ME as usual allowing for a transparent interpretation 
of the laws of thermodynamics. We note that a complementary analysis of quantum Otto cycles in the strong coupling 
regime, also employing the RC mapping, appears in a related work by Newman {\em et al.}~\cite{NewmanNazirPrelim}. Apart 
from the thermodynamic applications we propose here, it has been shown that such a
RC mapping and related concepts can provide a very accurate method to investigate the behaviour of open 
quantum systems for a variety of problems~\cite{JeanFriesnerFlemingJCP1992, MayKuhnSchreiberJPC1993, PollardFriesnerJCP1994, 
WolfsederDomckeCPL1996, WolfsederEtAlCP1998, HartmannGoychukHanggiJCP2000, ThossWangMillerJCP2001, 
MartinazzoHughesBurghardtPRE2011, IlesSmithLambertNazirPRA2014, BonfantiEtAlAnnPhys2015, WoodsPlenioJMP2016, 
IlesSmithEtAlJCP2016}. Even more generally, it is possible to apply this method iteratively by including several RCs
and in this way one can prove that every non-Markovian environment can be mapped to a Markovian
one~\cite{ChinEtAlJMP2010, MartinazzoEtAlJCP2011, WoodsEtAlJMP2014}.

Apart from adapting this general method to treat heat engines in the strong coupling and non-Markovian regime, we also
consider concrete applications. In particular, we find that non-Markovian effects can significantly enhance the steady
state efficiency even in the weak coupling regime; a result which -- to the best of our knowledge -- has not been found
before. In addition, we also investigate the relation between our method and the widely used polaron transformation,
showing that the RC framework provides a means to justify particular polaron transformed ME (PME), and reduces to it 
under special circumstances.

\emph{Outline: } We will start by introducing the general technique of the RC mapping in Sec.~\ref{sec RC mapping} as
far as it is needed to make the present treatment self-consistent. After having established this tool, we will present
the general thermodynamic framework based on it in Sec.~\ref{sec thermodynamics in general}. Particular applications
to devices working out of equilibrium then follow in Sec.~\ref{sec application 1} (efficiency study of a maser heat
engine in the non-Markovian regime) and Sec.~\ref{sec application 2} (a single electron transistor (strongly) coupled to
vibrations). Final remarks about the range of validity, open problems and the thermodynamic interpretation of the method
are given in Sec.~\ref{sec conclusions}.

\section{Reaction coordinate mapping}
\label{sec RC mapping}

We consider an arbitrary system with Hamiltonian $H_S(t)$ coupled linearly via some system operator $s$ to a bath
of harmonic oscillators (the coupling to several baths follows straightforwardly from this treatment, see next section).
The total Hamiltonian is assumed to have the typical Brownian motion form~\cite{WeissBook2008, GardinerZollerBook2004}
\begin{equation}\label{eq original BM Hamiltonian}
 H = H_S(t) + \frac{1}{2}\sum_k\left[p_k^2 + \omega_k^2\left(x_k - \frac{c_k}{\omega_k^2}s\right)^2\right]
\end{equation}
with mass-weighted positions $x_k$ and momenta $p_k$ of the bath fulfilling $[x_k,p_l] = i\delta_{kl}$ (we set
$\hbar\equiv1$ throughout). It is worth to point out that the completion of the square is important for a number of
reasons, e.g., to guarantee a thermodynamically stable Hamiltonian \emph{for all} coupling strengths
$c_k$~\cite{FordLewisOConnellPRA1988}. In the derivation of MEs one often neglects the quadratic system
``renormalization'' term $\frac{1}{2}\sum_k\frac{c_k^2}{\omega_k^2}s^2$ from the beginning, though its contribution
is, in principle, of the same order as the Lamb shift term.

An important result of the microscopic theory of Brownian motion is that the effect of the bath on
the system can be captured solely by one special function known as the spectral density (SD) of the bath:
\begin{equation}
 J_0(\omega) \equiv \frac{\pi}{2} \sum_k \frac{c_k^2}{\omega_k} \delta(\omega-\omega_k).
\end{equation}
The SD is a positive function for $\omega>0$ and must fulfill $J_0(\omega)\rightarrow0$ for $\omega\rightarrow0$
and $\omega\rightarrow\infty$.

\begin{figure}
 \includegraphics[width=0.39\textwidth,clip=true]{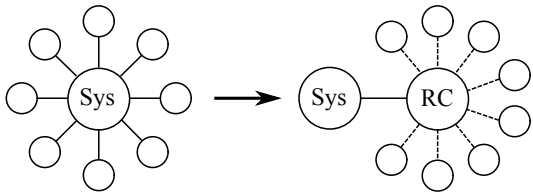}
 \label{fig RC mapping sketch 1}
 \caption{Sketch of the RC mapping. Before the mapping (left figure) the system can be visualized as being coupled to a
 large number of harmonic oscillators, see Eq.~(\ref{eq original BM Hamiltonian}). After the mapping (right figure) the
 system couples to the RC only, but in turn the RC is coupled to a large number of residual oscillators as described by
 the Hamiltonian~(\ref{eq Hamiltonian transformed BM form}). }
\end{figure}

Now, the spirit of the RC method is to define the interaction with the collective degrees of the freedom of the bath as 
one new coordinate: the reaction coordinate $X_1$ (see Fig.~\ref{fig RC mapping sketch 1}). Hence, we seek a 
transformation which maps 
\begin{equation}\label{eq def RC}
 \sum_k c_k x_k = \lambda_0X_1
\end{equation}
where $\lambda_0$ is an unspecified parameter so far. More formally, we perform a normal mode transformation of the
form
\begin{equation}
 \bb X = \Lambda\bb x, ~~~ \bb P = \Lambda\bb p.
\end{equation}
Here, we used a vector notation for the collection of original and transformed bath coordinates and momenta, e.g.,
$\bb x = (x_1,\dots,x_k,\dots,x_N)^T$ describes the original bath coordinates. For definiteness we considered a
finite number of $N$ bath oscillators (having the limit $N\rightarrow\infty$ in mind). Furthermore, $\Lambda$
is an orthogonal $N\times N$ matrix, i.e., $\Lambda^{-1} = \Lambda^T$, which guarantees that
$[X_k,P_l] = i\delta_{kl}$. Thus, $\Lambda$ has $\frac{N(N-1)}{2}$
independent components which are fixed by the requirement that the collection of residual bath oscillators (i.e., all
oscillators except the RC itself) is of normal form. This leads to
\begin{equation}
 \sum_k \omega_k^2 \Lambda_{lk}\Lambda_{mk} = \delta_{lm}\Omega_l^2
\end{equation}
for $l\neq1$ and $m\neq1$ and allows us to map the Hamiltonian~(\ref{eq original BM Hamiltonian}) to
\begin{equation}
 \begin{split}\label{eq Hamiltonian transformed 1}
  H'	=&~	H_S(t) + \frac{\delta\Omega_0^2}{2} s^2 - \lambda_0 s X_1 + \frac{1}{2}\left(P_1^2 + \Omega_1^2X_1^2\right)	\\
	&-	X_1 \sum_{k\neq1} C_kX_k + \frac{1}{2}\sum_{k\neq 1}\left(P_k^2 + \Omega_k^2X_k^2\right).
 \end{split}
\end{equation}
Here, we used $\Lambda_{1k} = \lambda_0^{-1}c_k$ which follows from Eq.~(\ref{eq def RC}). Furthermore, we defined
$\Omega_1^2 \equiv \sum_k\omega_k^2\Lambda_{1k}^2$ and $C_k \equiv -\sum_l \omega_l^2\Lambda_{kl}\Lambda_{k1}$.
Finally, the system gets renormalized due to $\delta\Omega_0^2 \equiv \sum_k \omega_k^{-2} c_k^2$ and from
$[X_1,P_1] = i$ we can deduce that $\lambda_0^2 = \sum_k c_k^2$.

At this point we can already recognize an important property of the mapping. Suppose that we scale the coupling
coefficients $c_k$ by $c_k \mapsto \alpha c_k$ for some $\alpha\in\mathbb{R}$. Then, the only parameters influenced by
this will be $\lambda_0$ and $\delta\Omega_0$, i.e., all information about the overall original system bath coupling
strength is captured in the system RC coupling and a system renormalization term. The remaining terms, especially the
new coupling coefficients $C_k$, are \emph{independent} of the initial coupling strength $\alpha$.

Now, the crux of the matter is that we do not have to determine $\Lambda$ directly; instead, the normal mode
transformation can be fully fixed by knowledge of the SD $J_0(\omega)$
only~\cite{GargOnuchicAmbegaokarJCP1985}. To see this we first of all note that all relevant quantities of the system
and RC itself can be expressed in terms of the original SD as follows:
\begin{align}
 \delta\Omega_0^2	&=	\frac{2}{\pi}\int_0^\infty d\omega \frac{J_0(\omega)}{\omega},	\label{eq help 1}	\\
 \lambda_0^2		&=	\frac{2}{\pi}\int_0^\infty d\omega \omega J_0(\omega),	\label{eq help 4}\\
 \Omega_1^2		&=	\frac{2}{\pi\lambda_0^2} \int_0^\infty d\omega \omega^3 J_0(\omega).	\label{eq help 2}
\end{align}
The effect of the residual environment on the system and RC is itself captured by the new SD
\begin{equation}
 J_1(\omega) \equiv \frac{\pi}{2} \sum_{k\neq1} \frac{C_k^2}{\Omega_k} \delta(\omega-\Omega_k)
\end{equation}
and the remaining task is to relate $J_0(\omega)$ to $J_1(\omega)$. We will here use the prescription given by Martinazzo
\emph{et al.}~\cite{MartinazzoEtAlJCP2011} who have shown the following relation under reasonable mild
conditions\footnote{$J_0(\omega)$ should be continuous and strictly positive for $\omega\in(0,\omega_R)$ and zero for
$\omega\ge\omega_R$ where $\omega_R$ denotes a cutoff frequency~\cite{MartinazzoEtAlJCP2011}. }
\begin{equation}\label{eq relation SD}
 J_1(\omega) = \frac{\lambda_0^2 J_0(\omega)}{|W_0^+(\omega)|^2}.
\end{equation}
For completeness we will present a derivation of this result in Appendix~\ref{sec appendix 1}. Similar methods to link
the SDs can be found in Refs.~\cite{GargOnuchicAmbegaokarJCP1985, ChinEtAlJMP2010, WoodsEtAlJMP2014,
IlesSmithLambertNazirPRA2014}. In Eq.~(\ref{eq relation SD}), $W_0(z)$ denotes the Cauchy transform of $J_0(\omega)$
given by
\begin{equation}
 W_0(z) \equiv \frac{1}{\pi} \int_{-\infty}^\infty d\omega\frac{J_0(\omega)}{\omega-z}
\end{equation}
[we extended $J_0(\omega)$ to negative values of $\omega$ via $J_0(-\omega) = -J_0(\omega)$] and we introduced the
notation
\begin{equation}
 W_0^+(\omega) \equiv \lim_{\epsilon\searrow0} W_0(\omega+i\epsilon) ~~~ (\omega\in\mathbb{R}).
\end{equation}
Furthermore, one can also show that (see Ref.~\cite{MartinazzoEtAlJCP2011} or Appendix~\ref{sec appendix 1} again)
\begin{equation}\label{eq relation RC frequencies}
 \frac{\lambda_0^2}{\delta\Omega_0^2} = \Omega_1^2 - \delta\Omega_1^2,
\end{equation}
where we denoted the frequency renormalization term of the RC by
\begin{equation}
 \delta\Omega_1^2 = \sum_{k\neq1} \frac{C_k^2}{\Omega_k^2} = \frac{2}{\pi}\int_0^\infty d\omega \frac{J_1(\omega)}{\omega}.
\end{equation}
Relation~(\ref{eq relation RC frequencies}) allows us to rewrite the Hamiltonian~(\ref{eq Hamiltonian transformed 1}) in
a Brownian motion form~\cite{BonfantiEtAlAnnPhys2015}
\begin{equation}
 \begin{split}\label{eq Hamiltonian transformed BM form}
  H'	=&~	H_S(t) + \frac{1}{2}\left[P_1^2 + \frac{\lambda_0^2}{\delta\Omega_0^2}\left(X_1 - \frac{\delta\Omega_0^2}{\lambda_0}s\right)^2\right]	\\
	&+	\frac{1}{2}\sum_{k\neq1}\left[P_k^2 + \Omega_k^2\left(X_k - \frac{C_k}{\Omega_k^2}X_1\right)^2\right],
 \end{split}
\end{equation}
which makes its thermodynamic stability evident. Hence, the \emph{physical} frequency of the RC is not given by
$\Omega_1$ but by the square-root of Eq.~(\ref{eq relation RC frequencies}).

Finally, we note that the power of the RC mapping also comes from the fact that it can be applied iteratively. This then
yields a chain of RCs where the last one is coupled to a residual environment. Remarkably, the relation between the
SDs, Eq.~(\ref{eq relation SD}), still carries over to this situation (replacing the index $0$ by $n$ and
the index $1$ by $n+1$ where $n$ labels the different RCs) and also all other parameters can be defined in terms of the
SD as in Eqs.~(\ref{eq help 1}) to~(\ref{eq help 2})~\cite{ChinEtAlJMP2010, MartinazzoEtAlJCP2011, 
WoodsEtAlJMP2014}. Furthermore and very importantly, the fixed point of this iteration scheme is a \emph{Markovian} 
SD~\cite{MartinazzoEtAlJCP2011} and the necessary conditions for convergence to a Markovian SD were worked out in 
Ref.~\cite{WoodsEtAlJMP2014} and are fulfilled for the situation considered here.\footnote{In the theory of Brownian 
motion, Markovian behaviour is ensured by an Ohmic SD which scales linearly with
$\omega$ up to a high enough frequency cutoff $\omega_R$ and then falls off to zero~\cite{WeissBook2008}. We stress,
however, that the correct definition of non-Markovianity in the quantum mechanical context is non-trivial, may not be 
guaranteed by this condition alone, and is under much
debate at the moment, see, e.g.,~\cite{WolfEtAlPRL2008, BreuerLainePiiloPRL2009, RivasHuelgaPlenioPRL2009}. } 
Thus, already at this time we can conclude 
that the dependence on the initial coupling strength is absorbed by including only \emph{one} RC 
(a more critical discussion of this point is shifted to Sec.~\ref{sec conclusions}) while strong
non-Markovianity might require several RCs.

\section{Nonequilibrium thermodynamics within the reaction coordinate framework}
\label{sec thermodynamics in general}

\begin{figure}
 \includegraphics[width=0.48\textwidth,clip=true]{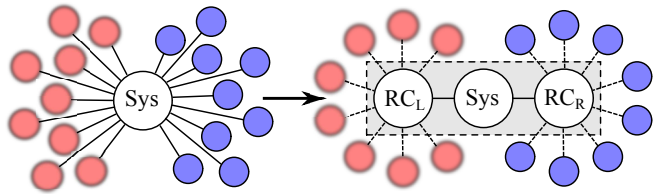}
 \label{fig RC mapping sketch 2}
 \caption{(color online). Sketch of a system coupled to a hot reservoir (red, blurred oscillators) and a cold reservoir
 (blue oscillators). After the mapping we have, as an example, included one RC for each reservoir to account for
 non-Markovian and strong coupling effects. Note, however, that we do not have to include a RC if the reservoir is
 weakly coupled and Markovian, or we might have to include two or more RCs in case of strong non-Markovianity. After the
 mapping we then treat the system and RCs as one new system as indicated by the shaded grey box. }
\end{figure}

We now imagine the situation where our system is coupled to several reservoirs labeled by $\nu$ and where the
time-dependent driving is responsible for work extraction and injection. The obvious generalisation of the
Hamiltonian~(\ref{eq original BM Hamiltonian}) to this situation is
\begin{equation}
 H = H_S(t) + \frac{1}{2}\sum_{k,\nu}\left[p_{k\nu}^2 + \omega_{k\nu}^2\left(x_{k\nu} - \frac{c_{k\nu}}{\omega_{k\nu}^2}s_\nu\right)^2\right]
\end{equation}
where the coupling to reservoir $\nu$ is mediated by the system operator $s_\nu$ which might be all different for
different $\nu$. A sketch of a possible scenario is shown in Fig.~\ref{fig RC mapping sketch 2}.
We can then decide to include zero, one or several RCs for each reservoir depending on the coupling strength and the
shape of the SD. Again, all what we need to know for this mapping are the SDs $J_0^{(\nu)}(\omega)$ for each reservoir
$\nu$. The same transformations as introduced in Sec.~\ref{sec RC mapping} will carry over in exactly the same way to
the situation of multiple reservoirs. It is in particular worth pointing out that each RC mapping is a unitary 
transformation \emph{only} on the Hilbert space of bath $\nu$, i.e., it leaves the system part and all other baths 
fully untouched. This feature allows us to really present a \emph{general} thermodynamic
framework valid for any system, which is coupled to its environment in the prescribed way.

After having included a sufficient number of RCs, the next step is to define a new ``supersystem'' consisting out of
the original system and all RCs. The idea is then to treat this supersystem within the standard Born-Markov-secular
(BMS) approximation, 
assuming that the bath of the residual oscillators is in thermal
equilibrium, and to derive a Markovian ME for the supersystem. This ME then has a transparent thermodynamic
interpretation (as we will review below for reasons of consistency) 
and this is indeed the strength of our approach: by finding this part of the environment which acts as an 
ideal, weakly coupled and memory-less thermal bath we are able to provide a formally clean definition of heat, which has 
a very precise meaning in thermodynamics and does \emph{not} simply equal the energy flowing into the surroundings in the 
general (i.e., non-Markovian and strongly coupled) case. Besides this fact, 
it is worth pointing out here that already
including one RC can give remarkable numerical results in agreement with the formally exact hierarchical equations of
motion method, as it was recently shown by Iles-Smith \emph{et al.}~\cite{IlesSmithLambertNazirPRA2014,
IlesSmithEtAlJCP2016}. Further research in this direction was also conducted in 
Refs.~\cite{JeanFriesnerFlemingJCP1992, MayKuhnSchreiberJPC1993, PollardFriesnerJCP1994, 
WolfsederDomckeCPL1996, WolfsederEtAlCP1998, HartmannGoychukHanggiJCP2000, ThossWangMillerJCP2001}. 

The standard framework of quantum thermodynamics starts with a microscopically derived ME of the
form~\cite{SpohnLebowitzAdvChemPhys1979, AlickiJPA1979, EspositoHarbolaMukamelRMP2009, KosloffEntropy2013,
SchallerBook2014, GelbwaserKlimovskyNiedenzuKurizkiAdv2015}
\begin{equation}\label{eq ME generic}
 \frac{d}{dt}\rho(t) = -i[H'_S(t),\rho(t)] + \sum_\nu \C L^{(\nu)}(t)\rho(t).
\end{equation}
Here, $\rho(t)$ denotes the density matrix and $H'_S(t)$ the Hamiltonian of the supersystem, i.e., the system \emph{and}
RCs. The time-dependence of $H'_S(t)$ might result from an initial time-dependence of $H_S(t)$. For
example, in case of a single RC we have [see Eq.~(\ref{eq Hamiltonian transformed BM form})]
\begin{equation}
 H'_S(t) = H_S(t) + \frac{1}{2}\left[P_1^2 + \frac{\lambda_0^2}{\delta\Omega_0^2}\left(X_1 - \frac{\delta\Omega_0^2}{\lambda_0}s\right)^2\right].
\end{equation}
Furthermore, the dissipators or thermal generators $\C L^{(\nu)}(t)$ are of Lindblad
form~\cite{BreuerPetruccioneBook2002} and fulfill local detailed balance, i.e.,
\begin{equation}
 \C L^{(\nu)}(t)e^{-\beta_\nu H'_S(t)} = 0
\end{equation}
where $\beta_\nu$ is the temperature of reservoir $\nu$ 
and which tells us that the ratio of backward to forward transition rates is given by a Boltzmann factor for 
the case of a Pauli rate ME for a non-degenerate supersystem Hamiltonian~\cite{SpohnLebowitzAdvChemPhys1979}.\footnote{If 
we also allow for particle transport by coupling
the system to a particle reservoir with chemical potential $\mu_\nu$, we have the relation
$\C L^{(\nu)}(t)e^{-\beta_\nu [H'_S(t)-\mu_\nu N'_S]} = 0$ instead where $N'_S$ is the particle number operator of the 
supersystem. } We remark that, strictly speaking, 
a ME of the form~(\ref{eq ME generic}) can 
only be derived for a slow time-dependence of $H_S(t)$. However, using techniques from Floquet theory it is also possible
to derive a ME for (strong) periodic driving~\cite{BreuerPetruccioneBook2002} with a similar thermodynamic
interpretation~\cite{KosloffEntropy2013, GelbwaserKlimovskyNiedenzuKurizkiAdv2015, BulnesCuetaraEngelEspositoNJP2015}.
For an arbitrary driving $H_S(t)$ there is no guarantee to find a simple ME for the system, but a thermodynamic
analysis can be still carried out~\cite{DeffnerLutzPRL2011} (see also the general
treatment~\cite{EspositoLindenbergVandenBroeckNJP2010}).

We now define the internal energy and entropy of the supersystem via
\begin{equation}
 E(t) \equiv \mbox{tr}\{H'_S(t)\rho(t)\}, ~~~ S(t) \equiv -\mbox{tr}\{\rho(t)\ln\rho(t)\}.
\end{equation}
The first law of thermodynamics then acquires the form
\begin{equation}
 \frac{d}{dt}E(t) = \dot W(t) + \sum_\nu\dot Q^{(\nu)}(t)
\end{equation}
where we identified the rate of work (power) \emph{done on} the supersystem and the heat flow coming \emph{from}
reservoir $\nu$ as\footnote{In presence of particle transport the heat flow
$\dot Q^{(\nu)}(t) = I_E^{(\nu)}(t) - \mu_\nu I_M^{(\nu)}(t)$ is composed of an energy current
$I_E^{(\nu)}(t) = \mbox{tr}\{H'_S(t)\C L^{(\nu)}(t)\rho(t)\}$ and a matter current
$I_M^{(\nu)}(t) = \mbox{tr}\{N'_S\C L^{(\nu)}(t)\rho(t)\}$ flowing into the supersystem. The first law then predicts
energy conservation, $\frac{d}{dt}E(t) = \dot W(t) + \sum_\nu I_E^{(\nu)}(t)$, and particle number conservation,
$\frac{d}{dt}N(t) = \sum_\nu I_M^{(\nu)}(t)$. }
\begin{align}
 \dot W(t)		&\equiv	\mbox{tr}\left\{\rho(t)\frac{d}{dt}H_S(t)\right\},	\label{eq def work flow}	\\
 \dot Q^{(\nu)}(t)	&\equiv	\mbox{tr}\left\{H'_S(t)\C L^{(\nu)}(t)\rho(t)\right\}.	\label{eq def heat flow}
\end{align}
In Eq.~(\ref{eq def work flow}) we used that $\frac{d}{dt}H'_S(t) = \frac{d}{dt}H_S(t)$.
The second law stipulates that the rate of entropy production $\dot S_\bb{i}(t)$ is always positive,
\begin{equation}\label{eq 2nd law}
 \dot S_\bb{i}(t) = \frac{d}{dt}S(t) - \sum_\nu \beta_\nu\dot Q^{(\nu)}(t) \ge 0.
\end{equation}
It is possible to prove Eq.~(\ref{eq 2nd law}) by use of Spohn's inequality stating that~\cite{SpohnJMP1978}
\begin{equation}\label{eq Spohns inequality}
 -\mbox{tr}\left\{\left[\C L^{(\nu)}(t)\rho(t)\right]\left[\ln\rho(t) - \ln \rho_\text{eq}^{(\nu)}\right]\right\} \ge 0
\end{equation}
for every $\nu$ and $\rho_\text{eq}^{(\nu)} \equiv e^{-\beta_\nu H'_S(t)}/Z_\nu$. By summing
Eq.~(\ref{eq Spohns inequality}) over $\nu$ we obtain the second law. 
At this point we wish to emphasize that within our theory the reservoirs $\nu$ enter additively 
(or separately) in the first and second law of thermodynamics, which is reminiscent of the fact that the RC mapping 
can be applied to each bath separately as indicated in Fig.~\ref{fig RC mapping sketch 2}. Especially, the temperatures 
(and chemical potentials) of the residual baths are still the same and well-defined. 

Finally, if the system is undriven it will eventually reach a steady state and the first and second law become
\begin{align}
 0		&=	\dot W + \sum_\nu \dot Q^{(\nu)},	\\
 \dot S_{\bb i}	&=	-\sum_\nu \beta_\nu\dot Q^{(\nu)} \ge 0.	\label{eq 2nd law steady state}
\end{align}
To indicate that we are at steady state, we dropped the time dependence on all quantities and for simplicity we will
exclusively focus on the steady state regime for the rest of this paper. We also note that even for an undriven system
there might be still a work source present (i.e., $\dot W\neq 0$) by identifying a work reservoir appropriately, see
Sec.~\ref{sec application 1}, or due to the possibility of particle transport (``chemical work''), see
Sec.~\ref{sec application 2}.

Before we proceed to illustrate our theory with examples of heat engines working out of equilibrium, it might be worth
to stress a simple consequence of our treatment at equilibrium. If the supersystem is time independent and in contact
with only one reservoir at inverse temperature $\beta$, it will relax to an equilibrium state
$\rho(t\rightarrow\infty) \sim e^{-\beta H'_S}$ such that the equilibrium state of the original system 
is~\cite{IlesSmithLambertNazirPRA2014} 
\begin{equation}\label{eq equilibrium state}
 \rho_S(t\rightarrow\infty) \sim \mbox{tr}_\text{RC}\{e^{-\beta H'_S}\}.
\end{equation}
In appendix~\ref{sec appendix 1 1} we will demonstrate that this state is consistent to lowest order 
perturbation theory in the coupling to the residual bath with the conventionally used Hamiltonian of mean 
force as introduced by Kirkwood~\cite{KirkwoodJCP1935}. In particular, 
this state in general does \emph{not} equal the canonical equilibrium state of the system alone, i.e.,
$\rho_S(t\rightarrow\infty) \not\sim e^{-\beta H_S}$. Experimentally, deviations from the canonical
state $e^{-\beta H_S}$ might be a clear indicator for persistent system environment correlations making it necessary to
go beyond the Born approximation, e.g., by using the RC mapping. Given the SD of the reservoir, it should be then
possible to test the prediction~(\ref{eq equilibrium state}) in an actual experiment.

\section{Application I: Three-level-maser heat engine}
\label{sec application 1}

\subsection{Standard treatment without RC}
\label{sec application 1 a}

\begin{figure}
 \includegraphics[width=0.30\textwidth,clip=true]{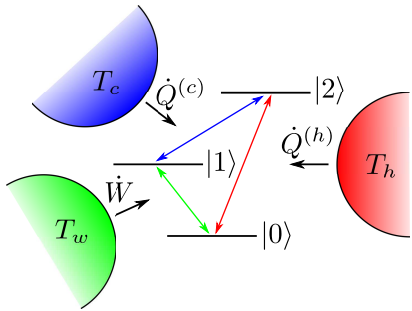}
 \label{fig maser heat engine}
 \caption{(color online). Sketch of the maser heat engine where the working medium comprises three discrete levels and
 each transition is coupled to a separate reservoir called the hot (``$h$'', red), cold (``$c$'', blue) and work
 (``$w$'', green) reservoir. The black arrows indicate the direction in which we define energy flows to be positive. }
\end{figure}

Possibly one of the simplest heat engine one can think of consists of three time-independent 
levels described by the Hamiltonian
\begin{equation}
 H_S = \sum_{i=0}^2 \epsilon_i |i\rangle\langle i|
\end{equation}
with $\epsilon_0<\epsilon_1<\epsilon_2$. The idea behind this engine is the model of a simple maser which is lasing at
a particular transition, say $0\leftrightarrow1$. This lasing corresponds to ``work'' output and it is achieved due to
population inversion between the levels $|0\rangle$ and $|1\rangle$. This in turn can be mediated via a third level
$|2\rangle$ due to the presence of two heat reservoirs at different temperatures (called the ``hot'' and ``cold''
reservoir respectively), also see Fig.~\ref{fig maser heat engine} for a sketch. Initially, this model was investigated
in 1959~\cite{ScovilSchulzDuBoisPRL1959}, but it is still of interest today~\cite{GevaKosloffJCP1996,
BoukobzaTannorPRL2005, CorreaEtAlSciRep2014, KosloffLevyAnnRevPC2014, CorreaEtAlPRE2014}.

The coupling to the reservoirs is mediated by the system operators
\begin{align}
 s_h	&=	\frac{|0\rl 2| + |2\rl 0|}{\sqrt{2\epsilon}},	\label{eq sh}	\\
 s_c	&=	\frac{|1\rl 2| + |2\rl 1|}{\sqrt{2\epsilon}},	\\
 s_w	&=	\frac{|0\rl 1| + |1\rl 0|}{\sqrt{2\epsilon}}.	\label{eq sw}
\end{align}
Here, in units where $\hbar=1$, the parameter $\epsilon$ has units of an energy or frequency such that $s_\nu$ has the
same units as the coordinates $x_{k,\nu}$ of the bath oscillators. However, in all numerical calculations which follow
we will simply set $\epsilon\equiv1$. Within the standard approach (BMS approximation for the system only) the thermal
generators $\C L^{(\nu)}$ in Eq.~(\ref{eq ME generic}) of each bath become
\begin{align}
 \C L^{(h)}	&=	\frac{J_0^{(h)}(\Delta_{20})}{\epsilon}\left\{[1+n_h(\Delta_{20})]\C D_{02} + n_h(\Delta_{20})\C D_{20}\right\},	\nonumber	\\
 \C L^{(c)}	&=	\frac{J_0^{(c)}(\Delta_{21})}{\epsilon}\left\{[1+n_c(\Delta_{21})]\C D_{12} + n_c(\Delta_{21})\C D_{21}\right\},	\nonumber 	\\
 \C L^{(w)}	&=	\frac{J_0^{(w)}(\Delta_{10})}{\epsilon}\left\{[1+n_w(\Delta_{10})]\C D_{01} + n_w(\Delta_{10})\C D_{10}\right\}.	\nonumber
\end{align}
Here, we have introduced the dissipator $\C D_{ij}\rho\equiv|i\rl j|\rho|j\rl i| - \frac{1}{2}\{|j\rl j|,\rho\}$, the
Bose distribution $n_\nu(\Delta_{ij}) \equiv (e^{\beta_\nu\Delta_{ij}} - 1)^{-1}$ and the SDs $J_0^{(\nu)}(\omega)$ of
bath $\nu$ are evaluated at the transition frequency $\Delta_{ij} \equiv \epsilon_i - \epsilon_j$.

Given the prescription of Sec.~\ref{sec thermodynamics in general}, it is not hard to compute the thermodynamic
behaviour of our system. At steady state the first law becomes $0 = \dot W + \dot Q^{h} + \dot Q^{c}$ with
$\dot W \equiv \dot Q^{w}$ while the second
law states that $-\beta_w \dot W -\beta_h\dot Q^{h} - \beta_c\dot Q^{c} \ge 0$. To quantify the performance of work
extraction (i.e., $\dot W < 0$), we introduce the efficiency of the heat engine:
\begin{equation}\label{eq def efficiency}
 \eta \equiv \frac{-\dot W}{\dot Q^{h}} = \frac{\Delta_{10}}{\Delta_{20}} \le \frac{\beta_c - \beta_h}{\beta_c - \beta_w}
\end{equation}
where the inequality is a consequence of the second law. Furthermore and very remarkably, it is possible to show that
the efficiency of the maser heat engine is always given by the ratio $\Delta_{10}/\Delta_{20}$ \emph{independent} of all
other parameters (as long as $\dot W < 0$, of course).

Note that to completely justify the notion of a ``work reservoir'', we should take the limit
$\beta_w\rightarrow0$~\cite{KosloffLevyAnnRevPC2014, CorreaEtAlPRE2014}, which complies with Sommerfelds notion of
temperature as the ``work value of heat''~\cite{SommerfeldBook1923}. In this limit the entropy change $-\beta_w\dot W$
in the work reservoir $w$ goes to zero, the second law acquires the form $-\beta_h\dot Q^{h} - \beta_c\dot Q^{c} \ge 0$
and the efficiency~(\ref{eq def efficiency}) is bounded by Carnot efficiency $\eta_\text{Carnot} = 1 - \frac{T_c}{T_h}$.
Microscopically, however, we recognize that the Bose distribution $n_w(\Delta_{10})$ diverges for
$\beta_w\rightarrow0$ (though $\dot W$ remains finite), unless we additionally require that the bath SD scales like
$J_0^{(w)}(\Delta_{10}) = \beta_w\Delta_{10}\Gamma_w$ for small $\beta_w$. In this ideal limit we then obtain
\begin{equation}
 \lim_{\beta_w\rightarrow0} \C L^{(w)} = \Gamma_w(\C D_{01} + \C D_{10}),
\end{equation}
i.e., the rates of upward and downward transitions are equal. However, for all numerical results reported below we take
$\beta_w$ to be small but non-zero.

Now, our approach is to consider the situation where it is actually not valid to apply this BMS ME for the system only. For
instance, this could be due to a structured (non-Markovian) SD. In this case, one should actually use a non-Markovian ME for the system
(e.g., the Redfield equation~\cite{BreuerPetruccioneBook2002}). However, the thermodynamic interpretation of
non-Markovian MEs is not clear and has not been established yet. In contrast, within our approach we know that we can
include a RC into our description in order to
account for non-Markovian effects on the three-level system (3LS) while the 3LS \emph{and} RC evolve in a
Markovian way. Any deviations from the efficiency $\Delta_{10}/\Delta_{20}$ then indicate non-Markovian and/or strong
coupling effects which we would be unable to detect by using the naive ME approach outlined in this section.

\subsection{Thermodynamics with RC}

\begin{figure}
 \includegraphics[width=0.47\textwidth,clip=true]{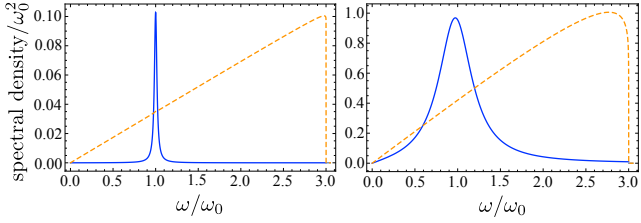}
 \label{fig plot spectral densities}
 \caption{(color online). Plot of the SDs $J_0^{(c)}(\omega)$ [Eq.~(\ref{eq SD J0c}), blue, solid] and
 $J_1^{(c)}(\omega)$ [determined by Eq.~(\ref{eq relation SD}), orange, dashed] over $\omega/\omega_0$ for two different
 values of $\gamma$ and $d_0$: $\gamma = 0.035\omega_0, d_0 = 0.06\omega_0$ (left) and
 $\gamma = 0.47\omega_0, d_0 = 0.67\omega_0$ (right) and $\omega_R = 3\omega_0$ in both cases.
 We recognize a pronounced peak at $\omega\approx\omega_0$ for small $\gamma$ in $J_0^{(c)}(\omega)$ whereas the shape
 of $J_1^{(c)}(\omega)$ remains rather unaffected. }
\end{figure}

For definiteness we choose the SD of the cold bath to be parameterized as
\begin{equation}\label{eq SD J0c}
 J_0^{(c)}(\omega) = \frac{d_0^2\gamma\omega}{(\omega^2-\omega_0^2)^2 + \gamma^2\omega^2} \Theta(\omega_R-\omega),
\end{equation}
while we still assume that it is safe to apply the Markov approximation with respect to the interaction with the hot
and work reservoir. Thus, by following the prescription of Sec.~\ref{sec RC mapping} we obtain the modified system
Hamiltonian
\begin{equation}
 H'_S = \sum_{i=0}^2 \epsilon_i |i\rangle\langle i| + \frac{1}{2}\left[P_1^2 + \frac{\lambda_0^2}{\delta\Omega_0^2}\left(X_1 - \frac{\delta\Omega_0^2}{\lambda_0}s_c\right)^2\right],
\end{equation}
and the SD $J_1^{(c)}(\omega)$ of the residual cold bath is determined by Eq.~(\ref{eq relation SD}). In
Fig.~\ref{fig plot spectral densities} we plot $J^{(c)}_0(\omega)$ and $J^{(c)}_1(\omega)$ for comparison.

At this point it is noteworthy that similar models have been also used in quantum biology to model light-harvesting
complexes~\cite{LambertEtAlNatPhys2013, HuelgaPlenioCP2013}. Indeed, guided by this motivation, a number of researchers
have started to investigate light-harvesting complexes from a heat engine perspective~\cite{KilloranHuelgaPlenioJCP2015,
XuWangZhaoCaoNJP2016} though we wish to stress that the models used there are not exactly the same as the one used
here. Furthermore, Ref.~\cite{KilloranHuelgaPlenioJCP2015} models the work reservoir effectively as a zero temperature
reservoir, which causes divergences in the standard thermodynamic formalism. In Ref.~\cite{XuWangZhaoCaoNJP2016} the
polaron transformation was used in order to access strong coupling effects. We will introduce this transformation in
Sec.~\ref{sec application 2} for a different model to compare it with our RC method.

To proceed we first of all note that the system Hamiltonian can be alternatively written as
\begin{equation}
 \begin{split}\label{eq Hamiltonian 3LS plus RC}
  &	H'_S = \left(\Delta_{10} + \frac{\delta\Omega_0^2}{4\epsilon}\right)|1\rl 1| + \left(\Delta_{20} + \frac{\delta\Omega_0^2}{4\epsilon}\right)|2\rl 2|	\\
  &	~~~~-\frac{\lambda_0}{\sqrt{2\epsilon}}X_1(|1\rl 2| + |2\rl1|) + \frac{1}{2}\left(P_1^2 + \frac{\lambda_0^2}{\delta\Omega_0^2}X_1^2\right),
 \end{split}
\end{equation}
where we made use of the fact that we can choose the ground state energy of the 3LS arbitrarily and set it to
$\epsilon_0 \equiv 0$. This Hamiltonian describes a Rabi model (harmonic oscillator coupled to a two-level system) plus
one additional energy level $|0\rangle$. Especially note that the energy levels of state $|1\rangle$ and $|2\rangle$ get
both shifted by the same amount $\delta\Omega_0^2/4\epsilon$. In the weak coupling (but non-Markovian) regime -- in
which we wish to compare our extended model with the one treated before -- these terms become negligible small.

\begin{figure}
 \includegraphics[width=0.43\textwidth,clip=true]{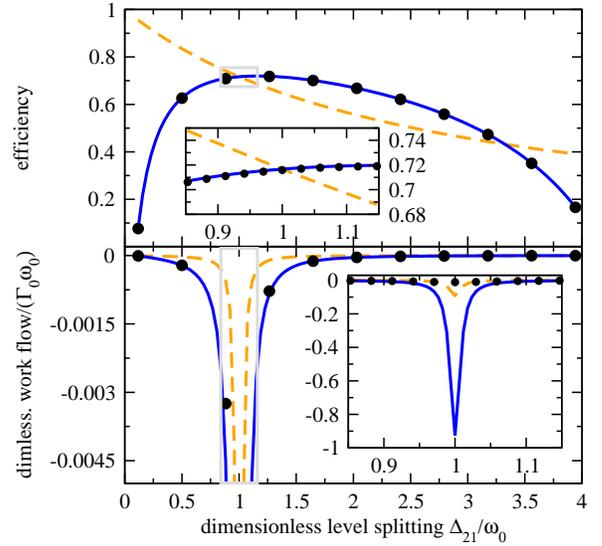}
 \label{fig plot efficiency work Delta20}
 \caption{(color online). Comparison of the efficiency (top) and the dimensionless work flow 
 $\dot W/(\Gamma_h\omega_0)$ (bottom) of the non-Markovian heat engine including the RC (BMS ME: blue solid lines; 
 ME without secular approximation: black dots) and the corresponding Markovian theory without RC from 
 Sec.~\ref{sec application 1 a} (orange dashed lines) as a function of the level splitting $\Delta_{21}/\omega_0$. 
 The insets zoom into the region indicated by the grey box in the main figues. 
 Parameters of the SD~(\ref{eq SD J0c}) are $d_0 = 0.0104\omega_0^2$, $\gamma = 0.0176\omega_0$, 
 $\omega_R = 588\omega_0$, and $\omega_0 = 0.17$. Furthermore, we chose $\Delta_{10} = 2.53\omega_0$, 
 $\beta_w\omega_0 = 0.0017$, $\beta_h\omega_0 = 0.17$ and $\beta_c\omega_0 = 17$ (implying a Carnot efficiency of 
 $\eta_\text{Carnot} = 0.99$). Finally, to completely specify the model we also need to fix the SDs of the hot and work 
 reservoir, which are determined by the coupling rates $\Gamma_w = 20\Gamma_h$ and $\Gamma_h = 0.001$, see 
 Appendix~\ref{sec appendix 2}. }
\end{figure}

To investigate the thermodynamic behaviour of our system, we will now use a ME which is explicit concerning the system-RC
interaction, but treats the coupling to the other reservoirs perturbatively and in a Markovian way. Following standard
procedures~\cite{EspositoHarbolaMukamelRMP2009, KosloffEntropy2013, SchallerBook2014, BreuerPetruccioneBook2002}
it is then possible to derive the BMS ME of the form~(\ref{eq ME generic}) with time-independent Hamiltonian $H'_S$
and dissipators $\C L^{(\nu)}$. Thus, the thermodynamic treatment follows straightforwardly from
Sec.~\ref{sec thermodynamics in general} and is formally equivalent to the model of Sec.~\ref{sec application 1 a}.
Hence, the first and second law become at steady state
\begin{align}
 0	&=	\dot{\tilde Q}^{h} + \dot{\tilde Q}^{c} + \dot{\tilde W},	\\
 0	&\le	-\beta_h\dot{\tilde Q}^{h} - \beta_c\dot{\tilde Q}^{c} - \beta_w\dot{\tilde W}.
\end{align}
Now, however, we used a tilde on all energy flows because they are numerically different from the corresponding
quantities in Sec.~\ref{sec application 1 a}. Likewise we introduce the efficiency of our engine for
$\dot{\tilde W} < 0$ as
\begin{equation}
 \tilde\eta = \frac{-\dot{\tilde W}}{\dot{\tilde Q}^{h}},
\end{equation}
which is also bounded by the Carnot efficiency in the limit $\beta_w\rightarrow0$.

Because it is not possible to diagonalize the Hamiltonian~(\ref{eq Hamiltonian 3LS plus RC}) in a simple manner, one has
to treat the ME numerically, and technical details of the derivation are presented in the Appendix~\ref{sec appendix 2}.
We also note that we decided to compare the ME based only on the Born and Markov approximation (i.e., \emph{without} secular
approximation) with the BMS ME. The latter is usually only well justified for systems where the level spacing is much
larger than the level broadening. This becomes increasingly questionable for more complex systems with many different
(and not always well separated) eigen frequencies. On the other hand, the advantage of the BMS ME is that the resulting
generator is of Lindblad form and allows for a mathematically clear proof of Spohn's
inequality~(\ref{eq Spohns inequality})~\cite{SpohnJMP1978, KosloffEntropy2013, BreuerPetruccioneBook2002}. In the
numerical results reported below, however, we never found any
violation of the second law even when we used the ME without secular approximation.

\begin{figure}
 \includegraphics[width=0.35\textwidth,clip=true]{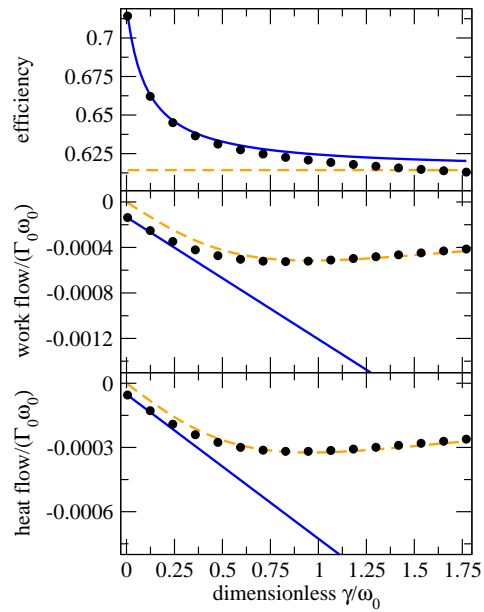}
 \label{fig plot efficiency work gamma}
 \caption{(color online). Comparison of the efficiency, the dimensionless work flow $\dot W/(\Gamma_h\omega_0)$ and 
 the dimensionless heat flow $\dot Q^{(c)}/(\Gamma_h\omega_0)$ into the cold reservoir of the non-Markovian heat engine 
 including the RC (BMS ME: blue solid lines; ME without secular approximation: black dots) and the corresponding 
 Markovian theory without RC from Sec.~\ref{sec application 1 a} (orange dashed lines) as a function of $\gamma$ [see 
 Eq.~(\ref{eq SD J0c})]. All parameters are as in Fig.~\ref{fig plot efficiency work Delta20} except that 
 $\Delta_{20} = 4.12\omega_0$. }
\end{figure}

\begin{figure}
 \includegraphics[width=0.48\textwidth,clip=true]{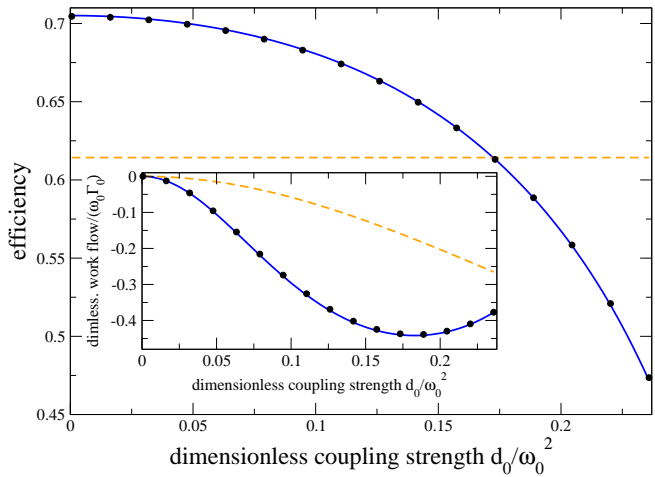}
 \label{fig plot efficiency work d0}
 \caption{(color online). Comparison of the efficiency and the dimensionless work flow $\dot W/(\Gamma_h\omega_0)$ 
 (inset) of the non-Markovian heat engine including the RC (BMS ME: blue solid lines; ME without secular approximation: 
 black dots) and the corresponding Markovian theory without RC from Sec.~\ref{sec application 1 a} (orange dashed lines) 
 as a function of the system-RC coupling strength $d_0/\omega_0^2$ (note that $d_0 = \lambda_0$ in our 
 case, see appendix~\ref{sec appendix 2}). All parameters are as in 
 Fig.~\ref{fig plot efficiency work Delta20} except that $\Delta_{20} = 4.12\omega_0$. }
\end{figure}

Figs.~\ref{fig plot efficiency work Delta20},~\ref{fig plot efficiency work gamma}
and~\ref{fig plot efficiency work d0} show numerical results obtained from the model without RC (dashed lines, see
Sec.~\ref{sec application 1 a}) and with RC [with (solid lines) and without (dots) secular approximation] for a specific
choice of parameters. This choice was done to illustrate -- from our point of view -- interesting features of Markovian
and non-Markovian heat engines. However, a detailed investigation of the model is beyond the scope of the paper and
would also be questionable because the model from Sec.~\ref{sec application 1 a} clearly is a simplified toy model.
Nevertheless, we wish to remark that the qualitative behaviour of our numerical results remains the same for a wide
range of parameters. Furthermore, we have focused on plots of the efficiency and work, which are both
of great interest in thermodynamics: obviously, we want to have a large power output, but on the other hand, a large
power output does not mean that our engine works more efficiently. In fact, if we possess a heat engine with low power
output but high efficiency, we can also build a machine with high power output and the same efficiency by running
several of these machines in parallel. Hence, we think efficiency is a more universal quantity on which one should put
more emphasis in the study of heat engines in the strong coupling and non-Markovian regime.

Turning to the details, we can infer from Fig.~\ref{fig plot efficiency work Delta20} two important points. First,
our approach predicts an efficiency enhancement of 10\% to 20\% in comparison to the standard theoretical framework
from Sec.~\ref{sec application 1 a}. We here note that it makes sense to compare the model without and with RC because
all parameters of the latter are completely fixed by the initial model. We simply use two different theoretical methods
to study the \emph{same} engine, but only the latter allows to capture, e.g., non-Markovian effects. The second point to
note concerns the secular approximation. Whereas for a large parameter regime it seems to be remarkably well justified,
it can also completely fail by predicting a nearly hundred times larger power output for the case
$\Delta_{21} \approx \omega_0$ (interestingly, this effect cancels when we compute the efficiency). In fact, if this
resonance condition is met\footnote{Note that $\omega_0$ equals the frequency of the RC for large cutoff frequency 
$\omega_R$, see appendix~\ref{sec appendix 2}. }, many energy levels are very close to each other (though never exactly 
degenerate) and a naive application of the secular approximation is not justified.

In Fig.~\ref{fig plot efficiency work gamma} we show the efficiency, work and heat flow as a function of $\gamma$.
As we can infer from Eq.~(\ref{eq SD J0c}) (also see Fig.~\ref{fig plot spectral densities}) a smaller $\gamma$ implies
a stronger peaked SD $J_0^{(c)}(\omega)$. Thus, $\gamma$ can be seen as a measure of the non-Markovianity of the SD
and Fig.~\ref{fig plot efficiency work gamma} proves that this is the cause of the efficiency enhancement. This
justifies our claim that non-Markovian machines can significantly outperform their Markovian
counterparts \emph{even at steady state}.\footnote{Note that Ref.~\cite{BylickaEtAlArXiv2015} also demonstrates an
increased ability to extract work due to non-Markovian effects. However, Ref.~\cite{BylickaEtAlArXiv2015} did not
study the efficiency and focused on transient effects, making it unclear whether a steadily working heat engine can
be better than its Markovian counterpart. } 
Physically, the reason for this can be traced back to the Purcell effect which predicts an enhanced 
spontaneous emission rate for the $1\leftrightarrow2$ transition in presence of a (resonant) cavity~\cite{PurcellPR1946} 
and thus allows for a stronger population inversion between the states $|0\rangle$ and $|1\rangle$. 
By including the RC we can indeed capture this 
effect which is clearly beyond the ``naive'' Markovian treatment of Sec.~\ref{sec application 1 a}. 
Furthermore, Fig.~\ref{fig plot efficiency work gamma} also shows that we
recover the results from Sec.~\ref{sec application 1 a} in the limit of large $\gamma$. In fact, for large $\gamma$
the SD does not only become more Markovian, but also the SD $J_1^{(c)}(\omega)$ of the residual cold bath is directly
proportional to $\gamma$, see Eq.~(\ref{eq SD J1c}), such that the RC evolves on time-scales much shorter then the 3LS
and can be adiabatically eliminated. Furthermore,
Fig.~\ref{fig plot efficiency work gamma} also shows that the BMS ME of the supersystem completely fails in this regime.
This behaviour can be traced back to the fact that the secular approximation does not commute with the adiabatic
elimination of the RC, i.e., the time-scales involved in the coherent evolution of the
system are of the same order as the dynamics of the relaxation due to the residual cold bath.

Finally,
Fig.~\ref{fig plot efficiency work d0} shows the thermodynamics as a function of the system-RC coupling strength $d_0$
(or the coupling strength between the 3LS and the cold reservoir, respectively). Again, we can observe a
strong efficiency enhancement and an almost perfect agreement between the secular and non-secular ME. Furthermore, we
observe that the efficiency decreases as a function of $d_0$ while the power output first increases up to a certain
critical coupling strength and then starts to decrease again (note that the power output for the 3LS ME
from Sec.~\ref{sec application 1 a} reaches a constant value for $d_0\rightarrow\infty$ instead). In fact, for all
parameters we have checked we always observed a decreasing efficiency as a function of $d_0$. Whether this is a general
feature or model specific remains an open question, which might be eventually answerable within the RC framework.

\section{Application II: Single electron transistor coupled to vibrations}
\label{sec application 2}

\subsection{Model and RC mapping}

As a second application we consider a single quantum dot in contact with two fermionic reservoirs (typically called a 
single electron transistor) and additionally coupled to a bath of phonons. 
Related models have been studied, e.g., in Refs.~\cite{KochVonOppenPRL2005, EntinWohlmanImryAharonyPRB2010,
MaierSchmidtKomnikPRB2011, WhiteGalperinPCCP2012, SchallerEtAlNJP2013, KrauseEtAlJCP2015}, in order to understand
electronic transport through molecules where the phonon bath models molecular vibrations. Here, as well as outlining 
another model applicable to the RC method, we also wish to compare with the polaron transformation, a technique 
frequently used to access the regime of strong system-phonon 
coupling~\cite{MaierSchmidtKomnikPRB2011, WhiteGalperinPCCP2012, SchallerEtAlNJP2013, KrauseEtAlJCP2015, 
NicolinSegalPRB2011, WangRenCaoSciRep2015, GelbwaserKlimovskyAspuruGuzikJPCL2015}. We will discuss what we mean by that 
in more detail at the end of Sec.~\ref{sec application 2 ME description}. 

The global Hamiltonian can be written as
\begin{equation}\label{eq Htot appl 2}
 H = H_\text{dot} + H_\text{el} + H_\text{el-dot} + H_\text{ph}.
\end{equation}
The electronic part of the system is described by
\begin{align}
 H_\text{dot}		&=	\epsilon d^\dagger d,	\\
 H_\text{el}		&=	\sum_{\nu}\sum_k \epsilon_{k\nu} c_{k\nu}^\dagger c_{k\nu},	\\
 H_\text{el-dot}	&=	\sum_{\nu}\sum_k (t_{k\nu}c_{k\nu} d^\dagger + t_{k\nu}^*d c_{k\nu}^\dagger)
\end{align}
where $d^{(\dagger)}$ and $c_{k,\nu}^{(\dagger)}$ are fermionic annihilation (creation) operators of the dot and the
reservoir with associated on-site energy $\epsilon$ and lead energy $\epsilon_{k\nu}$. Furthermore, the dot is connected
to two leads $\nu\in\{L,R\}$ with tunneling amplitudes $t_{k\nu}$. The interaction of the dot with the phonon bath is
assumed to be
\begin{equation}
 H_\text{ph} = \frac{1}{2}\sum_q \left[p_q^2 + \omega_q^2\left(x_q - \frac{h_q}{\omega_q^2} d^\dagger d\right)^2\right].
\end{equation}
Here, we denoted the coupling coefficients to the phonon bath by $h_q$ because we already use $c_{k\nu}$ for the
electrons in the fermionic reservoirs.

Next, again in order to overcome the limitations of the usual BMS ME we employ the RC mapping, this time to the phonon
bath. This transforms the system and phonon part to [see Eq.~(\ref{eq Hamiltonian transformed 1})]
\begin{align}
 & H_\text{dot} + H_\text{ph} =	\\
 & \left(\epsilon + \frac{1}{2}\sum_q\frac{h_q^2}{\omega_q^2}\right) d^\dagger d - \lambda_0 d^\dagger d X_1 + \frac{1}{2}\left(P_1^2 + \Omega_1^2 X_1^2\right)	\nonumber	\\
 & -X_1\sum_q C_q X_q + \frac{1}{2}\sum_q \left(P_q^2 + \Omega_q^2X_q^2\right).	\nonumber
\end{align}
We now identify our supersystem to be
\begin{equation}
 H'_S = \tilde\epsilon d^\dagger d - \lambda_0 d^\dagger d X_1 + \frac{1}{2}\left(P_1^2 + \Omega_1^2 X_1^2\right)
\end{equation}
with $\tilde\epsilon \equiv \epsilon + \frac{1}{2}\sum_q h_q^2/\omega_q^2$ whereas the interaction with the phonon bath
is described via $H^\text{ph}_I = -X_1\sum_q C_q X_q$ and the residual phonon bath Hamiltonian becomes
$H_B^\text{ph} = \frac{1}{2}\sum_q (P_q^2 + \Omega_q^2X_q^2)$. Note that this does not exactly correspond to the
mapping~(\ref{eq Hamiltonian transformed BM form}), but now the Hamiltonian is closer to the literature to which we
wish to compare our results~\cite{SchallerEtAlNJP2013}. Furthermore, the spectrum of $H'_S$ is still bounded from
below for all coupling strengths $\lambda_0$.

Below it will be convenient to work with the bosonic ladder operators, which are related to position and momentum
operators via
\begin{equation}
 X_q = \frac{1}{\sqrt{2\Omega_q}}\left(a_q + a_q^\dagger\right), ~~~ P_q = i\sqrt{\frac{\Omega_q}{2}}\left(a_q - a_q^\dagger\right).
\end{equation}
In terms of these operators we have
\begin{align}
 \tilde H_S	&=	\tilde\epsilon d^\dagger d + \Omega_1\left(a_1^\dagger a_1 + \frac{1}{2}\right) - \frac{\lambda_0 d^\dagger d}{\sqrt{2\Omega_1}} \left(a_1 + a_1^\dagger\right),	\label{eq HS appl 2}	\\
 H_I^\text{ph}	&=	-\left(a_1 + a_1^\dagger\right)\sum_q \frac{C_q}{2\sqrt{\Omega_1\Omega_q}}\left(a_q + a_q^\dagger\right),	\\
 H_B^\text{ph}	&=	\sum_q \Omega_q \left(a_q^\dagger a_q + \frac{1}{2}\right).
\end{align}

\subsection{ME description}
\label{sec application 2 ME description}

\begin{figure}
 \includegraphics[width=0.45\textwidth,clip=true]{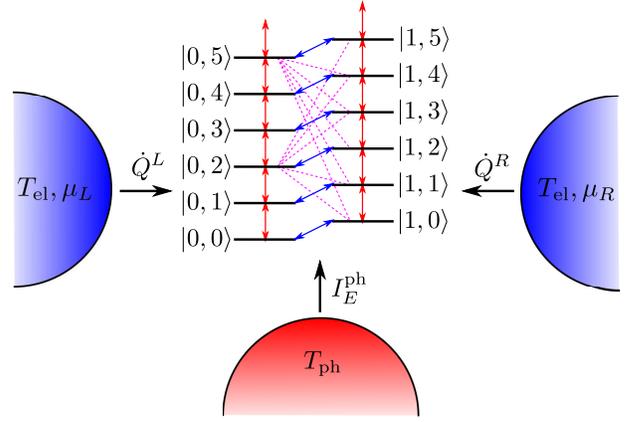}
 \label{fig sketch SET}
 \caption{(color online). This sketch shows the level structure of the Hamiltonian~(\ref{eq help 7}) with eigenstates
 $|n,m\rangle$. The phonon bath (red, vertical arrows) can induce transitions between states
 $|n,m\rangle \leftrightarrow |n,m+1\rangle$ whereas the electronic leads (blue, diagonal arrows) always change the
 occupation number $n$ of the quantum dot. Furthermore, the jump of an electron in or out of the system can be also
 accompanied by multi-phonon transitions as exemplarily indicated with thin, dashed, magenta lines. }
\end{figure}

To deduce the BMS ME we need to diagonalize the system Hamiltonian~(\ref{eq HS appl 2}). This can be
accomplished via the unitary transformation (we introduce $\lambda \equiv \lambda_0/\sqrt{2\Omega_1}$ for brevity)
\begin{equation}\label{eq polaron trafo}
 U = \exp\left[\frac{\lambda}{\Omega_1}d^\dagger d\left(a_1-a_1^\dagger\right)\right].
\end{equation}
It transforms operators according to $Ua_1 U^\dagger = a_1 + \frac{\lambda}{\Omega_1}d^\dagger d$ and
$UdU^\dagger = d e^{-\frac{\lambda}{\Omega_1}(a_1-a_1^\dagger)}$.  
Hence, applying~(\ref{eq polaron trafo}) to $H'_S$ yields
\begin{equation}\label{eq help 7}
 U H'_S U^\dagger = \bar\epsilon d^\dagger d + \Omega_1\left(a_1^\dagger a_1 + \frac{1}{2}\right)
\end{equation}
with $\bar\epsilon \equiv \tilde \epsilon - \lambda^2/\Omega_1$. This Hamiltonian is obviously in diagonal form with
eigenstates $|n,m\rangle$, where $n\in\{0,1\}$ ($m\in\{0,1,2,\dots\}$) quantifies the electronic (bosonic) occupation,
and eigenenergies $E_{nm} = \bar\epsilon n + \Omega_1(m+\frac{1}{2})$.

Given the spectral decomposition of the Hamiltonian it is simple to deduce the BMS ME, especially
if the spectrum is non-degenerate, which will generically be the case (unless $\bar\epsilon$ coincides with $\Omega_1$).
To completely specify our model we choose to define and parametrize the SD of the fermionic leads as
\begin{equation}\label{eq fermionic SD}
 \Gamma_\nu(\omega) \equiv 2\pi\sum_k |t_{k\nu}|^2\delta(\omega-\epsilon_{k\nu}) = \frac{\Gamma_\nu\delta_\nu^2}{(\omega-\varepsilon_\nu)^2 + \delta_\nu^2}.
\end{equation}
Here, $\delta_\nu$ quantifies the width and $\varepsilon_\nu$ the maximum of the Lorentzian function. Furthermore, the
SD of the residual phonon bath is choosen to be ohmic with exponential cutoff,
\begin{equation}
 J_1(\omega) = J_\text{ph} \omega e^{-|\omega|/\omega_R},
\end{equation}
which for large cutoff frequencies $\omega_R$ corresponds to an initial SD of the form~(\ref{eq SD J0c}) (as discussed
at the end of Appendix~\ref{sec appendix 2}). Further technical details concerning the derivation of the BMS ME are
provided in Appendix~\ref{sec appendix 3}. A sketch of the resulting dynamics is provided in Fig.~\ref{fig sketch SET}.

Before we proceed to show numerical results, let us briefly explain an alternative way to treat strong system-phonon 
interaction and to which we refer as the polaron ME (PME). Similarly to the RC mapping, also the PME starts with 
a unitary transformation, which acts on the bath \emph{and} system Hilbert space though. This transformation usually 
has the form of a generalized displacement operation such as~(\ref{eq polaron trafo}). Then, within this displaced 
reference frame it is possible to treat the coupling as a small perturbation again because the strong part is absorbed 
in this new frame. Consequently, it is possible to derive a ME valid for strong coupling for the \emph{system part only} 
(in our case here the quantum dot). In contrast, for our treatment we explicitly include the RC as a part of the (super) 
system and use the transformation~(\ref{eq polaron trafo}) later on only to formally diagonalize the supersystem.

\subsection{Results}
\label{sec application 2 results}

\begin{figure}
 \includegraphics[width=0.48\textwidth,clip=true]{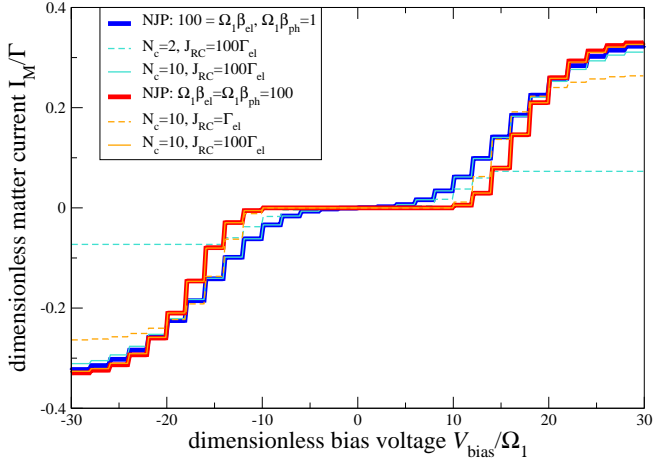}
 \label{fig comp_njp2013}
 \caption{(color online). Plot of matter current versus bias voltage for equal electronic tunneling rates 
 $\Gamma_L=\Gamma_R=\Gamma$. Previous results (bold curves, taken from~\cite{SchallerEtAlNJP2013}) are reproduced when 
 the coupling to the residual oscillators is much larger than to the electronic leads ($J_{\rm ph} \gg \Gamma$) and when 
 the cutoff $N_{\rm c}$ in the number of considered phonon modes is sufficiently 
 large. Other parameters have been chosen as $\delta_L=\delta_R=10\Omega_1$, $\varepsilon_L = \varepsilon_R=0$, 
 $\omega_{R}=10\Omega_1$, $\tilde\epsilon=0$, $\lambda_0=\sqrt{5}\Omega_1^{3/2}$. }
\end{figure}

In Fig.~\ref{fig comp_njp2013} we plot the matter current from left to right versus the difference in chemical 
potentials $V=\mu_L-\mu_R$ of the electronic reservoirs. 
First, we see that at low electronic temperatures, the current displays multiple steps, which occur at $V_i=2\Delta E_i$,
where $\Delta E_i$ are the transition frequencies of the system. Thus, the steps enable one to deduce the renormalized
electron energy $\bar\epsilon$ and the phonon frequency $\Omega_1$ from the electronic current. Furthermore, if we
truncate the phonon Hilbert space at a small cutoff number $N_{\rm c}$, we see that only few plateaus are visible since
the number of possible transitions is bounded. Most notably, however, when both $N_{\rm c}$ is large enough and the
coupling between the RC and the residual oscillators is large, we reproduce earlier results based on a
PME~\cite{SchallerEtAlNJP2013} (solid curves). In fact, for a large coupling between the RC and its environment, the RC
will thermalize on much shorter time-scales as compared to the dot-lead evolution. This is exactly the regime of the
PME in which it is assumed \emph{ad hoc} that the environment in the polaron frame is equilibrated with respect to the
dot state. We thus see that the RC method gives us a way to physically and mathematically justify the PME, but in
general will be also applicable beyond that regime.

Finally, we want to turn to the question whether it is possible to use our setup as a thermoelectric device, i.e.,
whether we can pump electrons against the bias due to a temperature gradient between electronic leads and phonon
reservoir. From Fig.~\ref{fig comp_njp2013} we see that we have zero current at zero bias even for different
temperatures of the electronic and phononic reservoirs (blue and turquoise). This is due to the fact that we chose
$\delta_L = \delta_R$ and $\varepsilon_L = \varepsilon_R$ [see Eq.~(\ref{eq fermionic SD})], which makes our setup 
symmetric under exchanging the 
labels $L$ and $R$ at zero bias. This makes thermoelectric transport impossible and will be changed now.

To quantify the irreversibility of our device we take a look at the entropy production~(\ref{eq 2nd law steady state}),
\begin{equation}
 \begin{split}
  \dot S_{\bb i}	&=	-\beta_L \dot Q^{L} - \beta_R \dot Q^{R} - \beta_\text{ph}\dot Q^{\text{ph}}	\\
			&=	\beta_\text{el}(\mu_L-\mu_R)I_M + (\beta_\text{el} - \beta_\text{ph})I_E \ge 0.
 \end{split}
\end{equation}
For the last line we assumed equal temperatures in the electronic leads $\beta_L=\beta_R=\beta_{\rm el}$ and we
used matter conservation $I_M \equiv +I_M^L = -I_M^R$ and energy conservation
$I_E \equiv I_E^{\rm ph} = -(I_E^L + I_E^R)$.

When the electronic and phononic temperatures are different (e.g. $\beta_{\rm el} > \beta_{\rm ph}$), heat will flow
between the electronic and phononic reservoirs, and the device can use a fraction of the heat to produce positive power
$P=-(\mu_L-\mu_R) I_M$ by transporting charges against the bias. The positivity of the entropy production leads to an
upper bound for the efficiency of heat-to-power conversion
\begin{equation}\label{eq efficiency appl 2}
 \eta = \frac{-(\mu_L-\mu_R) I_M}{I_E},
\end{equation}
which is defined for $(\mu_L-\mu_R) I_M < 0$.
As expected, one can show that the upper bound is given by Carnot efficiency.

\begin{figure}
 \includegraphics[width=0.48\textwidth,clip=true]{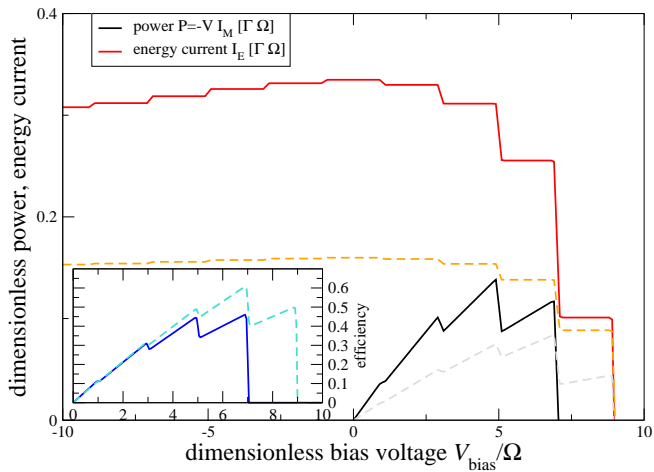}
 \label{fig efficiency}
 \caption{(color online).
 Plot of dimensionless power (solid black, dashed gray) and dimensionless energy current (solid red, dashed orange)
 versus bias voltage for equal coupling strengths to all reservoirs $\Gamma_L=\Gamma_R=\Gamma=J_{\rm ph}/2$. For
 $\delta_L=\delta_R=\Omega_1$ (solid curves) the efficiency (solid blue, inset) -- defined only in the region of
 positive power -- reaches about half Carnot efficiency at best. This is improved by narrowing the bandwidth to
 $\delta_L=\delta_R=0.5\Omega_1$ (dashed curves). Other parameters have been chosen as
 $\varepsilon_L=+5\Omega_1=-\varepsilon_R$, $\omega_R=10\Omega_1$, $\tilde\epsilon=0$, $\lambda_0=\Omega_1^{3/2}$,
 $\beta_{\rm el} \Omega_1=100$, $\beta_{\rm ph} \Omega_1=0.01$, $N_{\rm c}=10$. }
\end{figure}

To break the inherent left-right symmetry of our model, it is necessary to use different energy dependencies of the
electronic tunneling rates. We therefore consider the limit $\delta_L=\delta_R=\delta$ but
$\varepsilon_L\neq\varepsilon_R$. To make the effect rather strong we also have to consider $\delta$ smaller than
$|\varepsilon_\nu|$, but the use of the system as a thermoelectric device does not require that $J_{\rm ph} \gg \Gamma$.
In Fig.~\ref{fig efficiency} we plot the generated power and heat current from the hot phonon reservoir into the system
versus the bias voltage. First, we see that there exists a region where the generated power becomes positive, meaning
that charges are transported against the bias. This is only possible when heat flows from the phonon reservoir to the two
electronic reservoirs, and within the region of positive power we can see that the efficiency (inset) in
Eq.~(\ref{eq efficiency appl 2}) becomes finite and reaches for our chosen parameters about half Carnot efficiency. We
remark that a naive treatment of the quantum dot alone within a BMS ME would \emph{always} predict $I_E = 0$ because
$[H_\text{dot},H_\text{ph}] = 0$ and thus, no transport of electrons against the bias would be possible in this
framework. Having non-zero energy transport $I_E \neq 0$ is indeed a higher order effect (i.e., beyond second order
perturbation theory).

\section{Final remarks}
\label{sec conclusions}

We have presented a framework which allows to investigate thermal machines beyond the standard weak coupling and
Markovian assumption. This treatment is general in the sense that it can be applied to any system where strong
coupling effects and non-Markovianity are caused by a linear coupling with a bosonic bath. Our framework is by
construction thermodynamically consistent because we simply apply the standard framework of quantum thermodynamics to
an enlarged system. By this procedure we also automatically avoid spurious effects like efficiencies beyond Carnot.
Thus, in a certain sense the RC mapping helps us to find the correct border for which it is allowed to partition the
``universe'' into a ``system'' and a ``bath'' part, i.e., the partition for which it is justified to apply the Born
approximation.

Then, within this framework we have observed that non-Markovianity can indeed strongly enhance the efficiency of a heat
engine even in the weak coupling limit. Furthermore, we also observed that strong coupling decreases the efficiency,
though it is not clear at the moment whether this is a general feature. As another application we considered a
single electron transistor coupled to vibrations, which is an important model to understand molecular transport.
Besides studying its efficiency as a thermoelectric device we have demonstrated that the widely used PME
follows from our treatment as a special case and hence, justifying the PME.

Though we believe that we have introduced a very useful and practical framework, one should also critically question it. 
Especially, there might be (at first sight) two weak points on which we would like to comment.

First, though it was shown in Refs.~\cite{MartinazzoEtAlJCP2011, WoodsEtAlJMP2014} that the RC mapping is guaranteed to
give a quasi-Ohmic (Markovian) SD after we have included a sufficent number of RCs, it is not guaranteed that the 
resulting final SD is also weak in the sense that it is justified to consider only second order perturbation theory in 
the coupling to the residual baths. Though we have seen in Sec.~\ref{sec RC mapping} that the overall
coupling strength of the original system to the environment can be captured by using \emph{only one} RC, the resulting
final SD in general depends on the shape of the initial SD (but not on its ``absolute value'') and might be still large
compared to parameters of the system, which has to be checked in each case separately. Still, the RC method 
allows us to go beyond the standard perturbative approach in a reasonable way and furthermore, by applying a conceptually similar 
yet technically different mapping, one can show that also the resulting SD becomes small~\cite{HuhEtAlNJP2014}.  

Second, one might object that the information about what happens to the original system itself is somehow
lost. Indeed, it is true that we can only give expressions for the heat flows, the power and the entropy production for
the supersystem, including the RCs, but we do not know them for any particular subsystem. However, 
we have already stated in Sec.~\ref{sec thermodynamics in general} that the definition of heat becomes 
unambiguous only in case of a weakly coupled and memory-less thermal bath. In addition, we would like to defend this
approach by the following remarks. First of all, the division of a thermodynamic system into subsystems (e.g., system
and reservoirs) is in fact something which is only possible in the weak coupling limit. The very definition of ``system''
becomes ambiguous beyond that regime (also see the discussion on the zeroth law in~\cite{KosloffEntropy2013}).
Second, accessing the statistics of the system alone would require to incorporate explicit measurements 
in our framework, a problem which was so far completely ignored also in previous work on strong coupling and 
non-Markovian thermodynamics. In fact, even in the weak coupling regime one already relies on abstract tools such as 
full counting statistics to access the statistics of energy exchanges with a heat 
bath~\cite{EspositoHarbolaMukamelRMP2009}. However, work is usually regarded as a deterministic form of energy 
and as such should be easily accessible in experiments, for instance, by counting electrons as required for 
Sec.~\ref{sec application 2}. If the RC mapping is not directly applied to the ``work reservoir'', our framework 
is able to predict changed work statistics which should be measurable in real experiments. 
Third and finally, we point out that there has been progress to understand the energetics of arbitrary multipartite
systems \emph{locally}~\cite{WeimerEtAlEPL2008, HosseinNejadOReillyOlayaCastroNJP2015} and for special situations this
is also possible for the entropic balances~\cite{HartichBaratoSeifertJSM2014, HorowitzEspositoPRX2014} (note that the
meaning of ``bipartite'' in Refs.~\cite{HartichBaratoSeifertJSM2014, HorowitzEspositoPRX2014} is more restrictive than
in Refs.~\cite{WeimerEtAlEPL2008, HosseinNejadOReillyOlayaCastroNJP2015}). This opens up a new interpretation
of thermodynamics in the strong coupling and non-Markovian regime by recognizing the role played by the non-Markovian
environment as an effective feedback controller who acts back on the system based on the information stored about it.
Thus, advances in the thermodynamic understanding of multipartite systems will directly yield to new insights in the field
of strong coupling and non-Markovian thermodynamics.

\section*{Acknowledgments}

It is a pleasure to thank Javier Cerrillo, Mauro Cirio, Massimiliano Esposito, Jake Illes-Smith and Ahsan Nazir for 
valuable discussions as well as many participants of the fourth COST quantum thermodynamics conference in 
Erice. In addition, we explicitly wish to thank David Newman and Ahsan Nazir for sharing their related 
manuscript on thermodynamics and the reaction coordinate method prior to publication~\cite{NewmanNazirPrelim}. 
Furthermore, PS is especially indebted to Rocco Martinazzo for very helpful correspondence concerning the RC mapping.
Financial support of the DFG (SCHA 1646/3-1, SFB 910, and GRK 1558) is gratefully acknowledged. 
NL is partially supported by the FY2015 Incentive Research Project.


\bibliography{nazir,books,open_systems,thermo,info_thermo}

\appendix

\section{Mapping of the spectral densities}
\label{sec appendix 1}

We will here prove Eq.~(\ref{eq relation SD}) following the way of Martinazzo \emph{et al.}~\cite{MartinazzoEtAlJCP2011};
for other derivations see Refs.~\cite{GargOnuchicAmbegaokarJCP1985, ChinEtAlJMP2010, WoodsEtAlJMP2014,
IlesSmithLambertNazirPRA2014}. Because the system is completely arbitrary in our treatment,
we will choose for the moment without loss of generality a particle with position $q$ and momentum $p$ moving in a
potential $V(q)$ and coupled via the operator $s = q$ to the bath. The equations of motion according to 
the original Hamiltonian~(\ref{eq original BM Hamiltonian}) then take on the form
\begin{align}
 \ddot q		&=	-\frac{\partial V}{\partial q} + \sum_k c_k x_k - \sum_k \frac{c_k^2}{\omega_k^2}q,	\\
 \ddot x_k	&=	-\omega_k^2 x_k + c_k q.
\end{align}
After Fourier transformation according to the definition
\begin{equation}
 \hat f(z) \equiv \int_{-\infty}^\infty dt e^{izt} f(t) ~~~ (\Im(z)>0),
\end{equation}
we obtain
\begin{align}
 -z^2\hat q	&=	-\frac{\widehat{\partial V}}{\partial q} + \sum_k c_k\hat x_k - \sum_k \frac{c_k^2}{\omega_k^2} \hat q,	\\
 -z^2\hat x_k	&=	-\omega_k^2\hat x_k + c_k\hat q.
\end{align}
Eliminating $\hat x_k$ we can write
\begin{equation}
 -\frac{\widehat{\partial V}}{\partial q} = \hat L_0(z)\hat q.
\end{equation}
with the Fourier space propagator
\begin{equation}
 \begin{split}\label{eq help 3}
  \hat L_0(z)	&=	-z^2 - \sum_k \frac{c_k^2}{\omega_k^2-z^2} + \sum_k \frac{c_k^2}{\omega_k^2},	\\
		&\equiv	-z^2 - W_0(z) + \delta\Omega_0^2.
 \end{split}
\end{equation}
Here, we have introduced the Cauchy transform of $J_0(\omega)$:
\begin{equation}\label{eq def cauchy transform}
 W_0(z) \equiv \frac{2}{\pi}\int_0^\infty d\omega \frac{\omega J_0(\omega)}{\omega^2-z^2} = \frac{1}{\pi} \int_{-\infty}^\infty d\omega \frac{J_0(\omega)}{\omega-z}
\end{equation}
where we used $J_0(-\omega) = -J_0(\omega)$. As Leggett noticed~\cite{LeggettPRB1984}, the SD
of the bath is linked to the propagator via
\begin{equation}
 J_0(\omega) = -\lim_{\epsilon\searrow0} \Im[\hat L_0(\omega + i\epsilon)] \equiv -\Im[\hat L_0^+(\omega)] ~~~ (\omega\in\mathbb{R}).
\end{equation}
Especially within our notation we have $J_0(\omega) = \Im[W_0^+(\omega)]$ what can be directly proven by use of the
identity
\begin{equation}
 \delta(\omega'-\omega) = \lim_{\epsilon\searrow0} \frac{1}{\pi} \frac{\epsilon}{(\omega'-\omega)^2 + \epsilon^2}.
\end{equation}

For the next step we look at the transformed Hamiltonian~(\ref{eq Hamiltonian transformed 1}) to derive
\begin{align}
 \ddot q	&=	-\frac{\partial V}{\partial q} + \lambda_0 X_1 - \delta\Omega_0^2q,	\\
 \ddot X_1	&=	 - \Omega_1^2 X_1 + \lambda_0 q + \sum_k C_k X_k,	\\
 \ddot X_k	&=	 - \Omega_k^2 X_k + C_k X_1.
\end{align}
Playing the same game as above we can deduce that the Fourier space propagator for the system coordinate is
\begin{equation}
 \hat L_0(z) = -z^2 - \frac{\lambda_0^2}{\Omega_1^2 - z^2 - W_1(z)} + \delta\Omega_0^2
\end{equation}
which must be the same as Eq.~(\ref{eq help 3}). Furthermore, $W_1(z)$ is defined analogously to
Eq.~(\ref{eq def cauchy transform}) with $J_0(\omega)$ replaced by $J_1(\omega)$. Then, by comparison with
Eq.~(\ref{eq help 3}) we see that
\begin{equation}
 W_0(z) = \frac{\lambda_0^2}{\Omega_1^2 - z^2 - W_1(z)}.
\end{equation}
Rearranging terms we obtain
\begin{equation}
 W_1(z) = \Omega_1^2 - z^2 - \frac{\lambda_0^2}{W_0(z)}
\end{equation}
from which we can directly deduce the relation~(\ref{eq relation SD}). Furthermore, by noting that
$W_i(0) = \delta\Omega_i^2$ ($i=0,1$) we can also directly verify Eq.~(\ref{eq relation RC frequencies}).

\section{Non-canonical equilibriums states and the potential of mean force}
\label{sec appendix 1 1}

The Hamiltonian (or potential) of mean force is an elegant way to express the exact reduced system state 
of a thermal equilibrium system-bath state~\cite{KirkwoodJCP1935}, which was also used, e.g., in 
Refs.~\cite{JarzynskiJSM2004, CampisiTalknerHaenggiPRL2009, SeifertPRL2016}. The central idea is to 
introduce the Hamiltonian of mean force 
\begin{equation}
 H^* \equiv -\frac{1}{\beta}\ln\frac{\mbox{tr}_E\{e^{-\beta(H_S+H_I+H_E)}\}}{\mbox{tr}_E\{e^{-\beta H_E}\}} 
\end{equation}
such that the reduced system state can be expressed as 
\begin{equation}\label{eq eq state mean force}
 \rho_S = \frac{e^{-\beta H^*}}{Z^*_S} 
\end{equation}
with the partition function $Z^*_S = \mbox{tr}_S\{e^{-\beta H^*}\}$. Indeed, it is straightforward to show that 
Eq.~(\ref{eq eq state mean force}) coincides with the reduced equilibirum state of system \emph{and} environment, 
i.e., 
\begin{equation}
 \rho_S = \mbox{tr}_E\left\{\frac{e^{-\beta(H_S + H_I + H_E)}}{\mbox{tr}_{SE}\{e^{-\beta(H_S + H_I + H_E)}\}}\right\}.
\end{equation}
To see this it suffices to note that by definition 
$Z^*_S = \mbox{tr}_{SE}\{e^{-\beta(H_S + H_I + H_E)}\}/\mbox{tr}_E\{e^{-\beta H_E}\}$. 

To explore the connection of Eq.~(\ref{eq eq state mean force}) with the equilibrium state~(\ref{eq equilibrium state}) 
stated in the main text, we perform the RC mapping on the global system-bath Hamiltonian: 
\begin{equation}
 H_S + H_I + H_E = H'_S + H'_I + H'_E 
\end{equation}
with $H'_S = H_S + H_{S\text{-RC}} + H_\text{RC}$ describing the system, system-RC coupling and the RC respectively, 
and $H'_I$ describes the coupling to the residual bath described by $H'_E$. 
Since this mapping is exact, we can express the Hamiltonian of mean force (and consequently the partition function $Z_S$) 
in terms of the transformed Hamiltonian: 
\begin{equation}
 e^{-\beta H^*} = \frac{\mbox{tr}_{\text{RC},E'}\{e^{-\beta(H'_S + H'_I + H'_E)}\}}{\mbox{tr}_{\text{RC},E'}\{e^{-\beta(H_\text{RC} + H'_I + H'_E)}\}}. 
\end{equation}
Now, if it is justified to regard the coupling $H'_I$ to the residual bath as weak compared to all other contributions, 
we obtain to lowest (i.e., zeroth) order in $H_I$ the Hamiltonian of mean force and partition function 
\begin{equation}
 e^{-\beta H^*} \approx \frac{\mbox{tr}_{\text{RC}}\{e^{-\beta H'_S}\}}{\mbox{tr}_{\text{RC}}\{e^{-\beta H_\text{RC}}\}}, ~~~ Z^*_S \approx \frac{\mbox{tr}_{S,\text{RC}}\{e^{-\beta H'_S}\}}{\mbox{tr}_\text{RC}\{e^{-\beta H_{RC}}\}}. 
\end{equation}
Here, we used $[H_\text{RC},H'_E] = 0$, which is guaranteed by construction. Thus, to lowest order 
\begin{equation}
 \rho_S \approx \frac{\mbox{tr}_{\text{RC}}\{e^{-\beta H'_S}\}}{\mbox{tr}_{S,\text{RC}}\{e^{-\beta H'_S}\}} 
\end{equation}
as described by Eq.~(\ref{eq equilibrium state}) showing the consistency of our approach with standard results 
from equilibrium statistical mechanics. Especially, note that it is computationally relatively cheap to compute 
Eq.~(\ref{eq equilibrium state}) as compared to the exact result~(\ref{eq eq state mean force}).

\section{ME without secular approximation for application I}
\label{sec appendix 2}

We here provide details for the derivation of the Born-Markov ME based on the system
Hamiltonian $H'_S$~(\ref{eq Hamiltonian 3LS plus RC}), which treats the system-RC coupling non-perturbatively while
we are aiming at a perturbative treatment of the coupling to the remaining reservoirs. The coupling Hamiltonian is
\begin{equation}
 H_I = \sum_\nu H_I^{(\nu)} = \sum_\nu s_\nu\otimes B_\nu
\end{equation}
with $s_h$ and $s_w$ given in Eqs.~(\ref{eq sh}) and~(\ref{eq sw}) and $s_c = X_1$ due to the RC mapping. Furthermore,
for $\nu\in\{h,w\}$ the coupling operators of the bath are $B_\nu = -\sum_k c_{k\nu}x_{k\nu}$ and the free bath
Hamiltonian reads $H_B^{(\nu)} = \frac{1}{2}\sum_{k} (p_{k\nu}^2 + \omega_{k\nu}^2 x_{k\nu}^2)$. For $\nu = c$ the form
of the operators remains the same but we have to substitute $x_{kc}\rightarrow X_{kc}$, $p_{kc}\rightarrow P_{kc}$,
$c_{kc}\rightarrow C_{kc}$ and $\omega_{kc}\rightarrow \Omega_{kc}$.

Note that we are here neglecting renormalization terms of the form
$\frac{1}{2}\sum_k \frac{c_{k\nu}^2}{\omega_{k\nu}^2}s_\nu^2$ which are of second order in the system-bath coupling.
We will later on neglect any Lamb shift terms as well which are also of second order. This is consistent because
the heat flows~(\ref{eq def heat flow}) are itself already of second order in the coupling. The contribution due to the
renormalization and Lamb shift terms would then be of fourth order in total, which is beyond the validity of our
perturbative approach.

After applying the Born and Markov approximation, the formal starting point of the ME is the second order
equation in the interaction picture\footnote{The interaction picture is defined by
$\tilde A(t) \equiv e^{+i(H'_S+H_B)t} A e^{-i(H'_S+H_B)t}$. }~\cite{BreuerPetruccioneBook2002, KosloffEntropy2013,
SchallerBook2014}
\begin{equation}
 \begin{split}\label{eq ME start}
  \frac{d}{dt}\tilde\rho(t) = -\sum_\nu	&	\int_0^\infty d\tau \mbox{tr}_\nu\left\{\tilde H^{(\nu)}_I(t)\tilde H^{(\nu)}_I(t-\tau)\tilde\rho(t)R^{(\nu)}_0\right.	\\
					&	\left. - \tilde H^{(\nu)}_I(t)\tilde\rho(t)R^{(\nu)}_0\tilde H^{(\nu)}_I(t-\tau) + h.c.\right\}.
 \end{split}
\end{equation}
Here, $\tilde\rho(t)$ is the density matrix of the supersystem (in the interaction picture) and
$R^{(\nu)}_0 \sim e^{-\beta_\nu H^{(\nu)}_B}$ the equilibrium density matrix of the reservoir. Furthermore, the fact
that the ME additively decomposes into contributions from each reservoir $\nu$ is due to the fact that
$\mbox{tr}_\nu\{H_I^{(\nu)}R_0^{(\nu)}\} = 0$.

After introducing the bath correlation function
\begin{equation}
 C_\nu(\tau) \equiv \mbox{tr}_\nu\left\{\tilde B_\nu(\tau) B_\nu R^{(\nu)}_0\right\},
\end{equation}
the ME~(\ref{eq ME start}) can be expressed as
\begin{align}
 &	\frac{d}{dt}\tilde\rho(t) = -\sum_\nu\int_0^\infty d\tau	\\
 &	\times \left\{C_\nu(\tau)\left[\tilde s_\nu(t)\tilde s_\nu(t-\tau)\tilde\rho(t) - \tilde s_\nu(t-\tau)\tilde\rho(t)\tilde s_\nu(t)\right]\right.	\nonumber	\\
 &	~~~ \left. + C_\nu^*(\tau)\left[\tilde \rho(t)\tilde s_\nu(t-\tau)\tilde s_\nu(t) - \tilde s_\nu(t)\tilde\rho(t)\tilde s_\nu(t-\tau)\right]\right\}.	\nonumber
\end{align}
If we split the correlation function into real and imaginary part, $C_\nu(\tau) = \Re C_\nu(\tau) + i\Im C_\nu(\tau)$,
we can write
\begin{equation}
 \begin{split}\label{eq ME step 1}
  \frac{d}{dt}\tilde\rho(t)	=&	-\sum_\nu\int_0^\infty d\tau \Re C_\nu(\tau)\left[\tilde s_\nu(t),\left[\tilde s_\nu(t-\tau),\tilde\rho(t)\right]\right]	\\
				&	- i\sum_\nu\int_0^\infty d\tau \Im C_\nu(\tau) \left[\tilde s_\nu(t),\left\{\tilde s_\nu(t-\tau),\tilde\rho(t)\right\}\right].
 \end{split}
\end{equation}
Because
\begin{align}
 &	C_\nu(\tau)	\\
 &=	\sum_k \frac{c_{k\nu}^2}{2\omega_{k\nu}} \left\{[1+n_\nu(\omega_{k\nu})] e^{-i\omega_{k\nu}\tau} + n_\nu(\omega_{k\nu}) e^{i\omega_{k\nu}\tau}\right\}	\nonumber	\\
 &=	\frac{1}{\pi} \int_0^\infty d\omega J^{(\nu)}(\omega) \left\{[1+n_\nu(\omega)] e^{-i\omega\tau} + n_\nu(\omega) e^{i\omega\tau}\right\},	\nonumber
\end{align}
we obtain
\begin{align}
 \Re C_\nu(\tau)	&=	\frac{1}{\pi}\int_0^\infty d\omega J^{(\nu)}(\omega) \cos(\omega\tau)\coth\frac{\beta_\nu\omega}{2},	\\
 \Im C_\nu(\tau)	&=	-\frac{1}{\pi}\int_0^\infty d\omega J^{(\nu)}(\omega) \sin(\omega\tau).
\end{align}
Here, we used that $1+2n_\nu(\omega) = \coth\frac{\beta_\nu\omega}{2}$.
Finally, after leaving the interaction picture the ME~(\ref{eq ME step 1}) becomes
\begin{align}
 \frac{d}{dt}\rho(t)	=&	-i[H'_S,\rho(t)]	\\
			&	- \sum_\nu\int_0^\infty d\tau \Re C_\nu(\tau)\left[s_\nu,\left[\tilde s_\nu(-\tau),\rho(t)\right]\right]	\nonumber	\\
			&	- i\sum_\nu\int_0^\infty d\tau \Im C_\nu(\tau) \left[s_\nu,\left\{\tilde s_\nu(-\tau),\rho(t)\right\}\right].	\nonumber
\end{align}

This form is still not very useful for numerical implementation. To achieve this goal we write
\begin{equation}
 s_\nu(-\tau) = \sum_{kl} s_\nu^{kl} e^{-i\omega_{kl}\tau} |k\rl l|
\end{equation}
with $H'_S|k\rangle = E_k|k\rangle$ and $\omega_{kl} \equiv E_k - E_l$. Upon using the identity
\begin{equation}
 \int_0^\infty d\tau e^{\pm i\omega\tau} = \pi\delta(\omega) \pm i\C P\frac{1}{\omega},
\end{equation}
and neglecting the principal value (Lamb shift) terms,
we arrive at a ME of the form
\begin{equation}
 \begin{split}\label{eq ME final}
  &	\frac{d}{dt}\rho(t) = -i[H'_S,\rho(t)]	\\
  &	- \sum_\nu[s_\nu,[\chi_\nu,\rho(t)] + \sum_\nu[s_\nu,\{\Theta_\nu,\rho(t)\}].
 \end{split}
\end{equation}
The new operators appearing in this equation are given as
\begin{align}
 \chi_\nu 		&=	\frac{1}{2}\sum_{kl} J^{(\nu)}(\omega_{kl})\coth\frac{\beta_\nu\omega_{kl}}{2} s_\nu^{kl}|k\rl l|,	\\
 \Theta_\nu		&=	\frac{1}{2}\sum_{kl} J^{(\nu)}(\omega_{kl}) s_\nu^{kl} |k\rl l|.
\end{align}

Finally, the model is specified by choosing the SDs of the baths as discussed in the main text. For the work bath
we again use $J^{(w)}(\omega) = J_0^{(w)}(\omega) = \beta_w\Delta_{10}\Gamma_w$ to mimic a constant SD in the infinite
temperature limit. The SD of the hot bath is parameterized by
$J^{(h)}(\omega) = J_0^{(h)}(\omega) = \frac{\Gamma_h\omega}{\Delta_{20}}\Theta(\omega_R-\omega)$ and for the cold bath
$J^{(c)}(\omega) = J_1^{(c)}(\omega)$ is given by the relation~(\ref{eq relation SD}) with $J_0^{(c)}(\omega)$ given
in Eq.~(\ref{eq SD J0c}). We are especially interested in the regime of a large cutoff frequency
$\omega_R\gg1$ because this allows by virtue of the residue theorem to evaluate $J_1^{(c)}(\omega)$
exactly.\footnote{The residue theorem requires $J_0(\omega)$ to be analytic (except for
isolated poles), which is -- strictly speaking -- never the case for a hard
cutoff $\Theta(\omega_R-\omega)$. However, the discrepancy with the true solution vanishs for
$\omega_R\rightarrow\infty$. } Then, for $4\omega_0^2 > \gamma^2$, it follows that
\begin{equation}\label{eq SD J1c}
 J_1^{(c)}(\omega) = \gamma\omega\Theta(\omega_R-\omega).
\end{equation}
Furthermore, we also have $\lambda_0 = d_0$ and $\delta\Omega_0 = d_0/\omega_0$ such that the frequency of the RC
is $\frac{\lambda_0}{\delta\Omega_0} = \omega_0$.

The secular ME requires to perform an additional approximation on top of the Born-Markov ME~(\ref{eq ME final}).
This can be done by averaging the generator of the Born-Markov ME in the interaction picture in time (similar to a
rotating wave approximation)~\cite{EspositoHarbolaMukamelRMP2009, BreuerPetruccioneBook2002} or by dynamical coarse
graining of the time-evolution~\cite{SchallerBrandesPRA2008, SchallerBook2014}. We will skip any details here because
the secular ME is also reviewed in Appendix~\ref{sec appendix 3}.

\section{BMS ME for application II}
\label{sec appendix 3}

For a non-degenerate system Hamiltonian, it is well-known that the BMS ME yields a simple rate equation (``Pauli ME'')
in the eigenbasis of $H_S$~\cite{SpohnLebowitzAdvChemPhys1979, BreuerPetruccioneBook2002, SchallerBook2014}.
This can be put into the form
\begin{equation}\label{eq rate equation}
 \dot{P}_k = \sum_l W_{kl} P_l,
\end{equation}
where $P_k$ is the probability to find the system in state $|k\rangle$ and the transition rate from energy eigenstate
$l$ to $k$ is given by
\begin{equation}\label{eq transition rates def}
 W_{kl} = \sum_{\alpha\beta} \gamma_{\alpha\beta}(E_l-E_k) \braketmat{k}{A_\beta}{l} \braketmat{k}{A_\alpha^\dagger}{l}^*.
\end{equation}
Here, we assumed a general interaction Hamiltonian of the form $H_I = \sum_\alpha A_\alpha \otimes B_\alpha$.
Furthermore, the $\gamma_{\alpha\beta}$ denote the Fourier transforms of the reservoir correlation functions
\begin{equation}
 \gamma_{\alpha\beta}(\omega) = \int_{-\infty}^\infty d\tau \mbox{tr}_B\left\{B_\alpha(\tau) B_\beta R_0\right\} e^{+i\omega\tau}.
\end{equation}

For our model from Sec.~\ref{sec application 2}, we identify the coupling operators
\begin{align}
 A_1	&=	d,\qquad B_1 = \sum_{k} t_{kL} c_{kL}^\dagger,	\\
 A_2	&=	d^\dagger,\qquad B_2 = \sum_k t_{kL}^* c_{kL},	\\
 A_3	&=	d,\qquad B_3 = \sum_{k} t_{kR} c_{kR}^\dagger,	\\
 A_4	&=	d^\dagger,\qquad B_4 = \sum_k t_{kR}^* c_{kR},	\\
 A_5	&=	\frac{1}{\sqrt{2\Omega_1}}(a_1+a_1^\dagger),	\\
 B_5	&=	-\sum_{q} \frac{C_q}{\sqrt{2\Omega_q}} (a_q+a_q^\dagger),
\end{align}
and the non-vanishing correlation functions yield
\begin{align}
 \gamma_{12}(\omega)	&=	\Gamma_L(-\omega) f_L(-\omega),	\label{eq help 5}	\\
 \gamma_{21}(\omega)	&=	\Gamma_L(+\omega)[1-f_L(+\omega)],	\\
 \gamma_{34}(\omega)	&=	\Gamma_R(-\omega) f_R(-\omega),	\\
 \gamma_{43}(\omega)	&=	\Gamma_R(+\omega)[1-f_R(+\omega)],	\\
 \gamma_{55}(\omega)	&=	J_1(+\omega) [1+n_B(+\omega)].	\label{eq help 6}
\end{align}
Here, we have used the definition of the fermionic SD~(\ref{eq fermionic SD})
and the SD of the residual oscillators is as usual defined by
\begin{equation}
 J_1(\omega) = \frac{\pi}{2} \sum_{q} \frac{C_q^2}{\Omega_q} \delta(\omega-\Omega_q).
\end{equation}
From Eq.~(\ref{eq transition rates def}) and Eqs.~(\ref{eq help 5}) to~(\ref{eq help 6}) we see that the rates
additively decompose into left, right, and phonon contributions and the total rate matrix in
Eq.~(\ref{eq rate equation}) has the structure $W = W^{L} + W^{R} + W^{\text{ph}}$, where for $k\neq l$ we have
\begin{align}
 W_{kl}^{L}		=&~	\gamma_{12}(E_l-E_k) \abs{\braketmat{k}{d^\dagger}{l}}^2	\nonumber	\\
			&+	\gamma_{21}(E_l-E_k) \abs{\braketmat{k}{d}{l}}^2,	\\
 W_{kl}^{R}		=&~	\gamma_{34}(E_l-E_k) \abs{\braketmat{k}{d^\dagger}{l}}^2	\nonumber	\\
			&+	\gamma_{43}(E_l-E_k) \abs{\braketmat{k}{d}{l}}^2,	\\
 W_{kl}^\text{ph}	=&~	\gamma_{55}(E_l-E_k) \frac{1}{2\Omega_1} \abs{\braketmat{k}{(a_1+a_1^\dagger)}{l}}^2.
\end{align}
To make the transition rates explicit, we have to evaluate the matrix elements of the system coupling operators, too.
The electronic jumps can be separated into pure electronic transitions and bosonic excitations of the RC. Denoting by
$|\widetilde{n',m'}\rangle \equiv U|n,m\rangle$ the basis in the original frame, we obtain
\begin{align}
 \abs{\braketmat{n',m'}{A_1}{n,m}}^2	&=	\abs{\braketmat{\widetilde{n',m'}}{d e^{-\frac{\lambda}{\Omega_1}(a_1 - a_1^\dagger)}}{\widetilde{n,m}}}^2	\nonumber	\\
					&=	\delta_{n',0}\delta_{n,1} \abs{\braketmat{\widetilde{m'}}{e^{-\frac{\lambda}{\Omega_1}(a_1 - a_1^\dagger)}}{\widetilde{m}}}^2,	\\
 \abs{\braketmat{n',m'}{A_2}{n,m}}^2	&=	\delta_{n',1}\delta_{n,0} \abs{\braketmat{\widetilde{m'}}{e^{+\frac{\lambda}{\Omega_1}(a_1 - a_1^\dagger)}}{\widetilde{m}}}^2,
\end{align}
whereas the transitions triggered by the phonon reservoir simply yield
\begin{align}
 &	\abs{\braketmat{n',m'}{A_5}{n,m}}^2 = \frac{\delta_{n,n'}}{2\Omega_1} \abs{\braketmat{\widetilde{m'}}{\left(a_1+a_1^\dagger\right)}{\widetilde{m}}}^2	\nonumber	\\
 &	= \frac{\delta_{n,n'}}{2\Omega_1}\left[\delta_{m',m+1} (m+1) + \delta_{m',m-1} m\right].
\end{align}
Having the rates at hand, we are now finally able to compute the quantities shown in
Sec.~\ref{sec application 2 results}. A visualization of the resulting rate equation, which has a highly connected
structure, is also provided in Fig.~\ref{fig sketch SET}.

\end{document}